\newcommand{\f}[1]{Fig.~\ref{#1}}
\newcommand{\eq}[1]{Eq.~(\ref{#1})}
\newcommand{\eqs}[2]{Eqs.~(\ref{#1}) and~(\ref{#2})}
\def\be{\begin{equation}}
\def\ee{\end{equation}}
\def\ba{\begin{array}}
\def\ea{\end{array}}
  \newcommand{\beq}{\begin{equation}}    
  \newcommand{\eeq}{\end{equation}}      
\def\bea{\begin{eqnarray}}
\def\eea{\end{eqnarray}}
  \newcommand{\bc}{\begin{center}}       
  \newcommand{\ec}{\end{center}}         
  \newcommand{\bitem}{\begin{itemize}}   
  \newcommand{\eitem}{\end{itemize}}     
  \newcommand{\bnum}{\begin{enumerate}}  
  \newcommand{\enum}{\end{enumerate}}    
\def\ds{\displaystyle}

\def\d{\partial}
\def\l({\left(}
\def\r){\right)}

\documentstyle[prb,aps,multicol,graphicx,hhline,psfig]{revtex}
  \renewcommand{\narrowtext}{\begin{multicols}{2} \global\columnwidth20.5pc}
  \renewcommand{\widetext}{\end{multicols} \global\columnwidth42.5pc}

\def\nn{n}
\def\sgn{{\rm sgn}\,}
\def\tt{\tilde{t}}
\def\tx{\tilde{x}}
\def\dBadt{\dot{B_a}}

\def\IT{I}
\def\dIdt{\dot{I}}
\begin{document}

\title{Scaling and exact solutions for the flux creep problem
in a slab superconductor} 

\author{ D.~V. Shantsev, Y.~M.~Galperin, and T.~H.~Johansen}

\address{Department of Physics, University of Oslo, P. O. Box 1048, Blindern,
0316 Oslo, Norway}

\date{\today}

\maketitle

\begin{abstract}
The flux creep problem for a superconductor slab placed in 
a constant or time-dependent magnetic field is considered. Logarithmic dependence of
the activation energy on the current density is assumed, $U=U_0
\ln(J/J_c)$, with a field dependent $J_c$.  
The density $B$ of the magnetic flux penetrating into the superconductor,
is shown to obey a scaling law, i.e., the profiles $B(x)$
at different times $t$ can be scaled to a function of a single variable
$x/t^\beta$. We found exact solution for the scaling function 
in some specific cases, and an approximate solution for a general case.  
The scaling also holds for a slab carrying transport current $I$
resulting in a voltage $V \propto I^p$, where $p \sim 1$.
When the flux fronts moving from two sides of the slab collapse at the center, 
the scaling is broken and $V(I)$ crosses over to $V\propto I^{U_0/kT}$.
\end{abstract}

\bc
      { PACS 74.60.Ge, 74.25.Ha}
\ec

\narrowtext
\section{Introduction}

Thermally-activated hopping of flux lines between pinning sites, or
flux creep, controls both the magnetic and transport properties of 
superconductors under various external conditions. 
In particular, it is responsible for the frequently observed
fast magnetic relaxation and the highly nonlinear local 
voltage-current curves. Flux creep is specifically pronounced in   
high-temperature superconductors because there the pinning
energies are small, while the operation 
temperatures are high.\cite{giant,yeshurun}

In the majority of theoretical and experimental studies of flux 
creep\cite{yeshurun,GurKup,brandt-uni,Brandt}
a superconductor is first placed  
in an increasing or decreasing field, thus acquiring a non-zero
magnetization, $M$. The field is then kept constant and the
relaxation of $M$ with time is examined.   
In the present work we address a different problem -- flux penetration
into a non-magnetized superconductor, $M(t=0)=0$. 
After turning on 
a constant or time-dependent magnetic field, a flux front propagates 
from the surface with some time-dependent velocity. 
In the case of a long slab in parallel field two  planar flux fronts 
propagate from each side. 
Absence of a characteristic spatial scale suggests a possibility for scaling solutions, i.e., 
the flux density profile $B(x)$ at different times $t$ is a 
function of a single variable $x/t^\beta$, where $\beta$
is a constant. Indeed, such scaling was demonstrated in 
Ref.~\onlinecite{vinokur} for an instantaneous turn-on of
a constant applied field, and assuming a superconductor characterized by a
logarithmic dependence of the pinning energy on the current density.

In this work we have sought for the whole class of
flux creep problems having solutions with a single-parameter scaling. 
This class turned out to be a lot broader than considered in 
Ref.~\onlinecite{vinokur}, insofar as it extends 
to situations with time-dependent applied fields, $B_a(t) \propto t^\alpha$, 
and a general field-dependent critical current density $J_c(B)$.
We also report scaling solutions for the flux creep in a superconductor 
carrying a transport current. There the flux creep
manifests itself in a number of experimental observations
like a relaxation in the resistance, and that the voltage across the sample 
depends on the current sweep rate, etc.\cite{zhang-ic,zhang,zhang99,ma,zeng} 

The paper is organized as follows.
In Sec.~II the basic equations are formulated.
Section~III brings out the scaling properties of their solution. 
In Sec.~IV two cases allowing an exact analytical solution are considered. 
The subsequent Sections report the implications of our results
for measurable quantities - $B$, $J$, $E$ distributions,
magnetization and voltage. Finally, the main conclusions are summarized.  
 
\begin{figure} 
\centerline{\psfig{figure=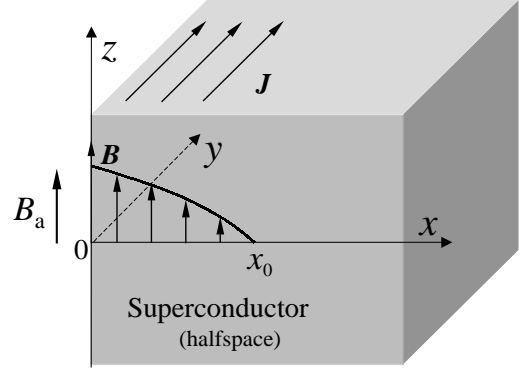,width=8.1cm}}
\caption{Superconductor slab in an applied magnetic field 
\label{f_slab}}
\end{figure}

\section{Formulation of the problem}

Consider a semi-infinite superconductor (filling the half-space $x>0$)
placed in an external magnetic field $B_a(t)$ directed along the  
$z$-axis, see \f{f_slab}. The position- and time-dependent flux density 
(or magnetic induction) and shielding current in the sample are denoted 
as  $B(x,t)$ and $J(x,t)$, respectively. Both $\bf J$ and the electric 
field $\bf E$ are directed along the $y$ axis. 

{}From the Ampere and Faraday laws one then has, 
\begin{eqnarray}
\mu_0 J &=& - \partial B / \partial x  \, , \label{J}\\
\partial B / \partial t &=& -  \partial E /\partial x \, .
\label{dBdt}
\end{eqnarray}
To describe the superconductor we assume 
that $E=vB$, where $v=v_0 \exp(-U/kT)$ is the velocity of the
thermally-activated vortex motion over the barrier $U$. 
With a logarithmic dependence $U(J)=U_0 \ln(J_c/J)$ it follows that
\beq
    E(J,B) = v_0 |B| \; \left| J/J_c\right|^\nn  \sgn J \, ,
\label{EBJ}
\eeq
where the exponent $\nn = U_0/k_BT$ depends on the temperature $T$.
The critical current density $J_c$ usually depends on
$|B|$, and in the bulk part of the paper it will be assumed 
that this is described by,\cite{Irie}
\be
  J_c(B) = J_{c0}\, |B_0/B|^{\gamma} \, .
\label{jcb}  
\ee
To avoid parameters with inconvenient dimensions we introduced here
two constants,  $J_{c0}$  and  $B_0$, a characteristic current
density and induction, in addition to the exponent $\gamma$.\cite{2param} 
With $\gamma=0$ one has a constant $J_c$ (the Bean model), while 
for $\gamma=-1/n$ one has an $E-J$ relation, 
\beq
    E = v_0 B_0 \,\left| J/J_{c0}\right|^\nn \, \sgn J\, , 
\label{EJ}
\eeq
which is $B$-independent. Such an $E(J)$ has been used by many authors 
since it allows a significant simplification of the analysis.
It is justified when the magnetic field inside  the superconductor
is essentially uniform, in particular, when a small perturbation 
$\delta B(x,t)$ is added to a superconductor cooled in a large dc field. 

The electrodynamic problem defined by Eqs.~(\ref{J})-(\ref{jcb}) can be
reformulated as a non-linear diffusion equation for the flux density,
\beq
    \frac{\partial B}{\partial t} = \frac{v_0}{(\mu_0 J_{c0} B_0^\gamma)^\nn} \
    \frac{\partial }{\partial x}  \;
    \l( |B|^{1+\gamma n}
     \left| \frac{\partial B} {\partial x} \right|^{\nn-1} \, 
     \frac{\d B}{\d x}
    \r).
\label{B}    
\eeq
Similar equations can be written also for $J$ and $E$.

We will first consider a totally flux free superconductor 
which at $t \geq 0$ experiences an increasing applied magnetic field given by
\be
B_a(t) = B_0\,(t/\tau)^\alpha \,   \, ,\quad  \alpha>0 \, ,  
\ee  
where $\tau$ is another constant. For $\alpha=0$ and $\alpha=1$ this describes
an instant field step and a linear ramping up, respectively. 
By introducing the dimensionless variables 
\begin{equation}
 b=\frac{B}{B_0}, \   \tilde{x} = x\, \frac{ \mu_0 J_{c0}}{B_0},  \ \tilde{t} =
 \frac{t}{\tau} ,  \ j = \frac{J}{J_{c0}} ,  \ \epsilon = \frac{E}{v_0
 B_0} \, ,
 \label{norm}
\end{equation}
and removing the redundant parameter definition by choosing 
\be
B_0 = \mu_0 J_{c0} v_0 \tau \, ,
\label{tau}  
\ee
the \eq{B} acquires the form\cite{specialcase}
\beq
    \frac{\partial b}{\partial \tt} = 
    \frac{\partial }{\partial \tx}  \;
    \l( |b|^{1+\gamma n}
     \left| \frac{\partial b} {\partial \tx} \right|^{n-1} \, 
     \frac{\d b}{\d \tx}
    \r) \, .
\label{b}    
\eeq
The boundary condition becomes 
\be
  b(0,\tilde{t}) = \tt^\alpha \, , 
\label{boundary1}
\ee
at the superconductor surface.

\section{Scaling}

The flux creep problem \eqs{b}{boundary1} is solved by writing the
flux density in the scaling form 
\beq
b(\tilde{x}, \tilde{t}) = \tt^\alpha f(\xi) \, , \quad \xi =
\tilde{x} \, \tilde{t}^{-\beta} \, ,\\ 
\label{sc}
\eeq
with
\beq
\beta=\frac{1+\alpha n (1+\gamma)}{1+n}
\label{beta}
\eeq
By substitution one finds that the scaling function $f(\xi)$ satisfies
the differential equation
\be
- \alpha f + \beta \xi f' = \l( f^{1+\gamma n} \, |f'|^{n} \r)'\,
 , \label{f}
\ee
where it was used that $f \ge 0$, and $f' \le 0$, i.e., the flux density 
decreases monotonously as one moves away from the surface. The boundary conditions become 
\be
 f(0) = 1\, , \quad f(\xi_0)=0\, , \label{boundary}
\ee
where the last one arises from the physical requirement that
the scaling function vanishes at the flux front, which is located at 
$\xi =\xi_0$.
{}From \eq{sc} it follows immediately that the flux front advances with time 
according to 
\be
  \tx_0(\tt) = \xi_0\, \tt^\beta \, .
\label{ff}  
\ee
For the case $\alpha=0$ (constant applied field) one has $\beta=1/(n+1)$, while
a larger $\alpha$ increases the exponent $\beta$.

Using \eqs{J}{EBJ} one finds that also the current density and electric field have
scaling properties, namely
\begin{eqnarray}
 &&j(\tx,\tt) =  \tt^{\alpha-\beta}\ |f'(\xi)| \,, \nonumber \\ 
&&\epsilon(\tx,\tt) = \tt^
  {\,\alpha+\beta-1}
  \ f_\epsilon(\xi), \quad
  f_\epsilon \equiv |f'|^n \ f^{1+\gamma n} \, .\label{jesc}
\end{eqnarray}
The scaling behavior of the ac losses is found from a similar analysis 
of the product $j(\tx,\tt)\epsilon(\tx,\tt)$.

When the magnetic field is instantly turned on and then kept constant, 
$(\alpha =0)$, exact scaling relations are obtained even when the power-law 
\eq{jcb} is relaxed and replaced by {\em any} $J_c(B)$. 
Considering here 
\be
J_c(B) = \frac{J_{c0}}{ s (B/B_0)} \, , 
\ee
where $s$ is a general function, a diffusion equation for the flux density 
similar to \eq{b} can be derived giving,
\be
    \frac{\partial b}{\partial \tt} = 
    \frac{\partial }{\partial \tx}  \,
    \left[ |b|\, s^n(b)
     \left| \frac{\partial b} {\partial \tx} \right|^{n-1} \, 
     \frac{\d b}{\d \tx}
    \right] \, ,
\label{bkim}    
\ee
where the same dimensionless variables \eqs{norm}{tau} are used. 
The scaling form of the flux density (\ref{sc}) is still applicable where now $b(\tx,\tt)=f(\tx \, \tt^{-1/(n+1)})$. 
The equation for the scaling function $f(\xi)$ is changed to
\be
  \xi f'/(n+1) = \left[ f \; s^n(f) \; |f'|^{n} \right]'
\label{fkim}  
\ee
while the boundary conditions (\ref{boundary}) are the same. 
The flux front, current density and electric field are now given by
\begin{eqnarray}
 &&  \tx_0 = \xi_0\ \tt^{~1/(n+1)}\, , \nonumber \\
&&j(\tx,\tt) =  \tt^{-1/(n+1)}\ |f'(\xi)| \, ,\nonumber \\
&& \epsilon(\tx,\tt) = \tt^{-n/(n+1)} f_\epsilon(\xi), \quad
  f_\epsilon \equiv |f'|^n\ f\ s^n(f)\, ,
\end{eqnarray}
hence, also they possess scaling properties in this case.

\begin{figure}
\centerline{\psfig{figure=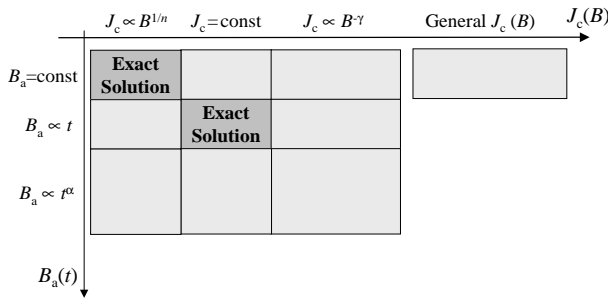,width=8.1cm}}
\vspace{0.3cm}
\caption{Summary of our results. The diagram illustrates where the flux
creep problem has scaling solutions. Full analytical solution is found 
for two particular cases. 
\label{f_pd}}
\end{figure}

As we have formulated the flux creep problem, the wide variety of
possible situations is represented by the set of three independent 
parameters, $n$, $\gamma$ and $\alpha$.
For a given parameter set the scaling function $f(\xi)$ can
be obtained by solving the \eqs{f}{fkim} numerically. Interestingly,
one may in two special cases find exact analytical solutions, as 
shown in the next Section. Shown in \f{f_pd} is a summary of our 
findings regarding scaling solutions for the flux creep 
problem in various situations. From this 
``$J_c(B)-B_a(t)$'' diagram it is seen that scaling 
holds for a majority of conceivable conditions.  
Note that even if the scaling function itself is unknown one may
test predicted scaling behaviors experimentally, e.~g.,
by analyzing flux density profiles $B(x,t)$ measured at different times.
Plotting $B(x,t)/B_a(t)$
versus $x/x_0(t)$ is expected to result in a collapse into a scaling function $f(\xi)$.
Other practical consequences of the scaling concerning the magnetization
and voltage behavior are discussed below.

\section{Exact solution for particular cases}

\subsection{Constant field, B-independent E(J)\label{exactcon}}

The solution for this case will be worked out in considerable detail
in order to illustrate some basic features of the scaling properties
of the physical quantities. 
Suppose that at $t=0$ a constant applied magnetic field, $B_0$, is  
turned on,
i.~e., one has $\alpha=0$. Furthermore, let $\gamma=-1/n$, which leads 
to the commonly assumed current-voltage law, \eq{EJ}. 
{}From \eqs{sc}{beta} the scaling law has then the form
\beq
b(\tx,\tt) = f\l(\xi \r), \quad \xi = x \, \tt^{-1/(n+1)}\, ,
\label{scconst}
\eeq
and  \eq{f} for the scaling function reduces to
\beq
\left[ \xi - \nn (\nn +1) \, |f'|^{\nn -2} \, f''  \right] \; f' = 0 \ .
\label{eqf} 
\eeq
This equation has two real solutions for  $f'$ consistent with
 $ \partial B / \partial x \leq 0$, namely
\beq
-f' = \left\{ \begin{array}{ll}
 \ds{
\left[ \frac{\nn -1}{2\nn(\nn +1)}\, \left( \xi_0^2 - \xi^2 \right)
\right]}^{1/({\nn-1})} 
 \! \!, & \xi \leq \xi_0 \, , \\[4mm]
0 \ , & \xi \geq \xi_0 \, .
            \end{array} \right.
\label{fd}
\eeq
The parameter $\xi_0$ is an integration constant which is
determined from the following interpretation of \eq{fd}. 
Noting that since $- f'(\xi)$ directly represents the spatial profile of the current $J(x,t)$, the solution with $f'= 0$ is identified as the one valid for the Meissner-state region where also $ b=0$. The
point $\xi = \xi_0$, then defining the flux penetration front, is 
determined by $f(\xi_0) =0$ using the solution valid for
the penetrated (mixed-state) region. We find
\beq
    \xi_0 = \left[
    2\nn \frac{\nn +1}{\nn -1} \; F(1)^{1-\nn}
    \right]^{{1}/({\nn +1})} \ ,
\eeq
where the function $F$ is defined as
\beq
    F(z) = \int_0^z (1-y^2)^{{1}/({\nn -1})} \, dy \ .
\label{F}    
\eeq
The value $F(1)$ can be expressed through $\Gamma$-functions as
\be
 F(1) 
= \frac{\sqrt{\pi}}{2} \ \Gamma\l(\frac{n}{n-1} \r) \, / \;
 \Gamma\l(\frac32 +\ds\frac 1{n-1} \r) \, . 
\ee
The result for the flux density profile becomes
\beq
    b(\tilde{x}, \tilde{t}) = 1 - 
    \frac{F({\tilde{x}}/{ \tilde{x}_0(\tilde{t})})}{F(1)} \, ,
 \quad   0 \leq \tilde{x} \leq \tilde{x}_0 \, ,
\label{bF}  
\eeq
where the position of the flux front, $\tilde{x}_0$, is moving with time according to
\beq
     \tilde{x}_0(\tilde{t}) = \xi_0 \, \tilde{t}^{~{1}/(\nn +1)} \ .
\label{x0}
\eeq

It then also follows that the normalized current density 
$ j = - \partial b/ \partial \tilde{x}$, and electric field in the flux penetrated region, $0 \leq \tilde{x} \leq \tilde{x}_0$, are given by
\begin{eqnarray}
  &&j(\tilde{x}, \tilde{t}) = \ds{ \frac{1}{F(1)} \,
    \frac{1}{\tilde{x}_0(\tilde{t})} \,
    \left[ 1 - \l( \frac{\tilde{x}}{\tilde{x}_0(\tilde{t})} \r)^2
    \right]^{{1}/({\nn -1})} }\! , \nonumber \\
  && \epsilon(\tx,\tt)=j(\tx,\tt)^n\, .  
\end{eqnarray}
The differential resistivity $\d E/\d J \propto j^{\nn -1}$ therefore
varies in space as a parabola having a maximum value at the edge.

\begin{figure} 
\vbox{
\centerline{\psfig{figure=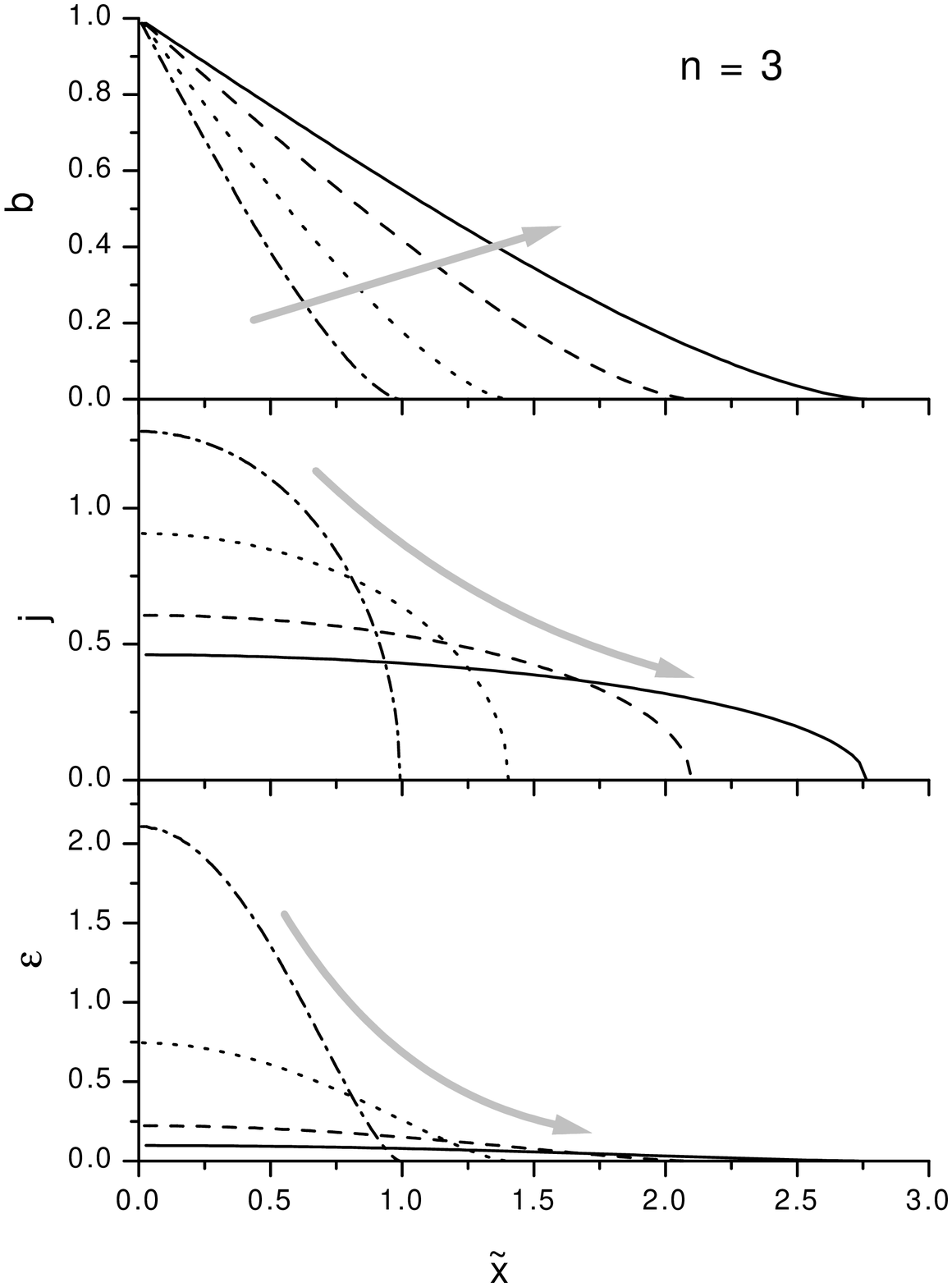,width=8.1cm}\psfig{figure=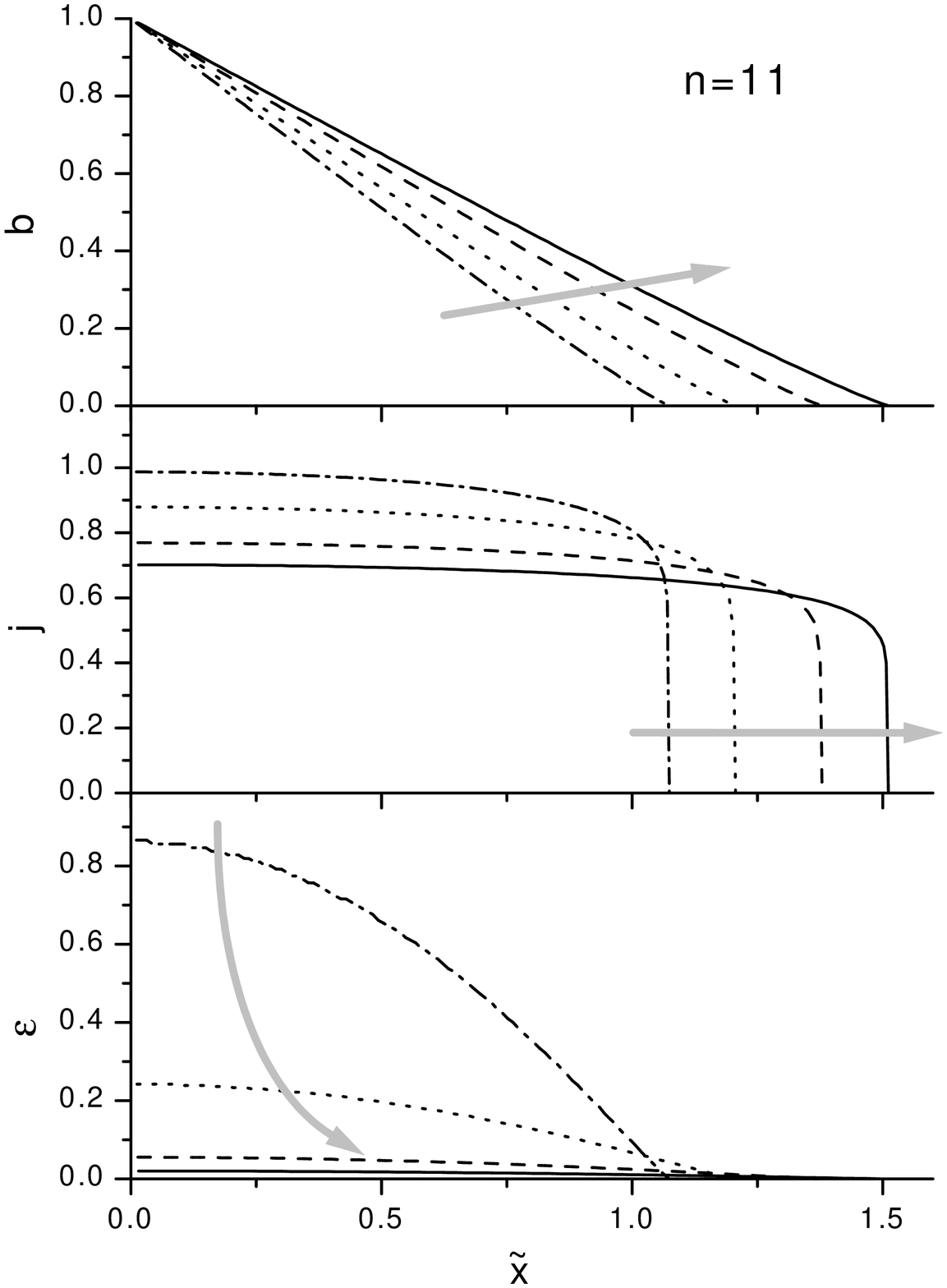,width=8.1cm}}
\caption{Distributions of the normalized flux density, current density
 and electric field after a sudden turn-on of a constant applied magnetic 
field. A $B$-independent $E(J)$, i.e., $\gamma=-1/n$, is assumed, and the plots are made for $n=3$ and 11. The arrows indicate time direction, and the curves correspond to $\tt=$0.05, 0.2, 1, and 3. 
\label{f_exact2}}}
\end{figure}

Figure~\ref{f_exact2} shows the distribution of the magnetic induction, 
current density and electric field for  $\nn =3$ and 11. 
As the time increases, the flux penetrates deeper into the sample 
accompanied by a smaller slope of the flux density profile, i.~e., 
a decay of current density. 
The electric field profile follows the behavior of $J(x)$, although
decreasing much more rapidly due to the
strongly nonlinear $E(J)$ law. As the power $\nn$ increases the field and current distributions are seen to become more linear. 
In the limit $\nn  = \infty$ one has $F(z)=z$ and $\xi_0 =1$, and hence
 $ \tilde{x}_0(\tilde{t}) =1 $. The temporal dependences then vanish,
and the behavior reduces to  $ b=  1-x/\delta$, $j=1$, and
$\epsilon=0$ for $0 \leq x \leq \delta$, where $\delta=B_0/\mu_0 J_{c0}$. 
Thus, the well-known results of the Bean model are reproduced. 

\widetext
\narrowtext

\subsection{Linearly increasing field, constant $J_c$\label{exactlin}}

For the linear sweep case, $B_a(t) = \dBadt t$ ($\dBadt=\,$const), and a constant $J_c$,  i.~e., $\alpha=1$ and $\gamma=0$, the flux creep problem has a surprisingly simple solution. It follows from \eq{B} that
\be
  B(x,t) = \dBadt t\, [ 1-x/x_0(t)], \quad x \le x_0 \, ,
\label{t-x}
\ee
and the region $x>x_0$ is flux free. The flux front propagates 
linearly in time, 
\be
  x_0(t)=\frac{\dBadt}{\mu_0 J_c} \l(\frac{\mu_0 J_c
    v_0}{\dBadt}\r)^{1/(n+1)} \, t \, . 
\label{t-xx0}  
\ee
Remarkably, this solution coincides with the Bean model behavior
where in the penetrated region the flux profile is a straight line, and
the current density is uniform,
\be
J(x) =  J_c \l(\dBadt/\mu_0 J_c v_0\r)^{1/(n+1)}=
\text{const}\, .
\label{jconst}
\ee
The electric field $E(x)$ becomes proportional to $B(x)$,
\be
  E(x,t) = v_0 \l(\dBadt/\mu_0 J_c v_0\r)^{n/(n+1)}\, B(x,t) \, ,
\label{eexactlin}  
\ee
hence, also this quantity is linear in both $x$ and $t$.
All these profiles are presented in \f{f_a1}(b).
For larger field sweep rates, $\dBadt$, the current density 
and electric field are higher,
while the penetration depth at a given $B_a$ is smaller.

The results of this subsection, unlike the results of the rest of the paper,
can be used also  to describe the flux distribution 
at large enough fields when the flux penetrates the whole slab.
The current then flows in the opposite
directions in the left and right halves of the slab, but with the same density
as defined by \eq{jconst}. 
The magnetic field in the left half is given by a linear profile defined by 
\eq{t-x}, where $x_0$ is  still given by \eq{t-xx0}, though it does 
not have the meaning of the flux front position anymore.
In the right half $B(x)$ is a mirror image of that in the left half.

\section{Summary of $B$, $J$, and $E$}

\begin{figure} 
\centerline{\psfig{figure=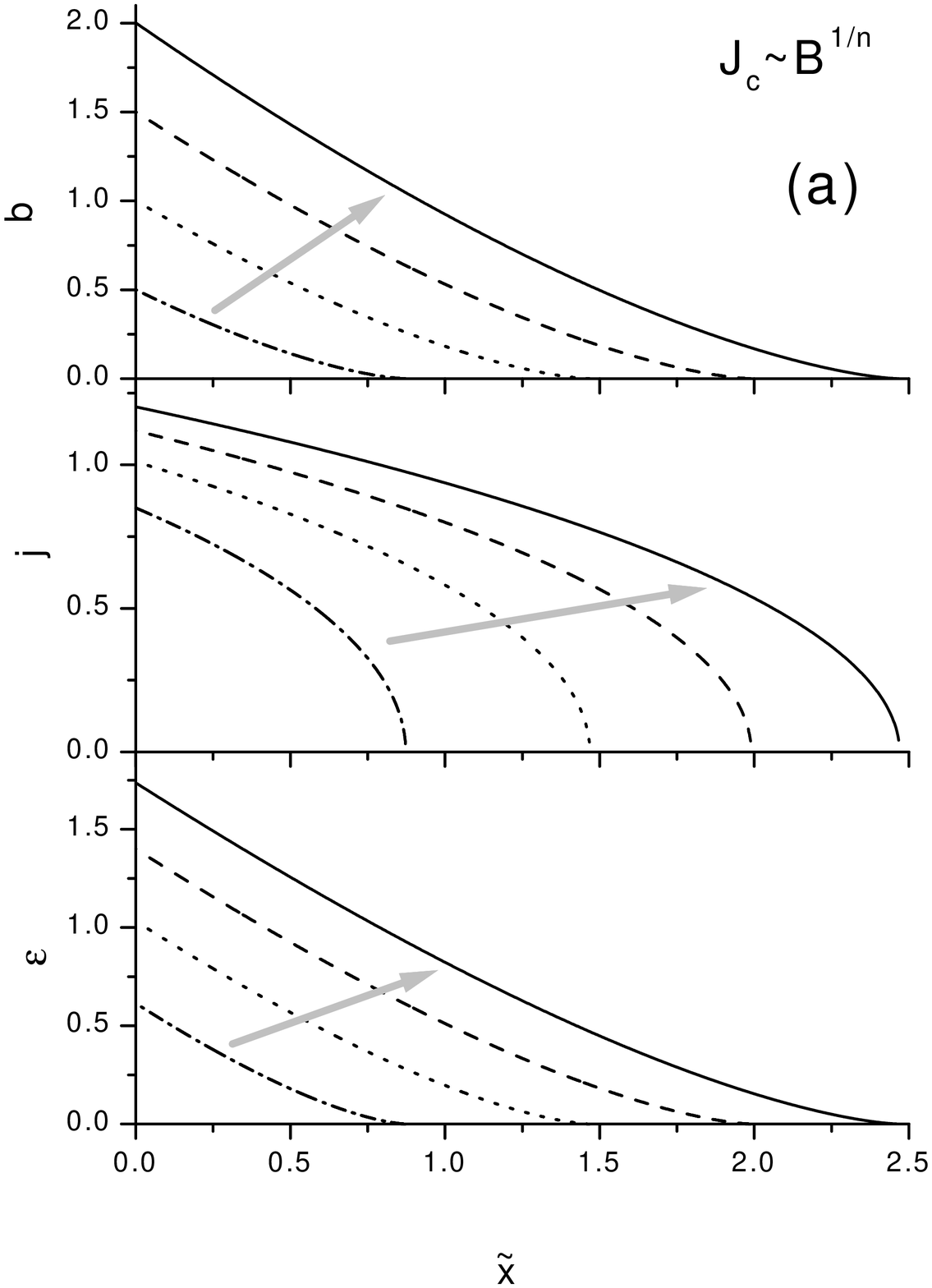,width=7.5cm}\psfig{figure=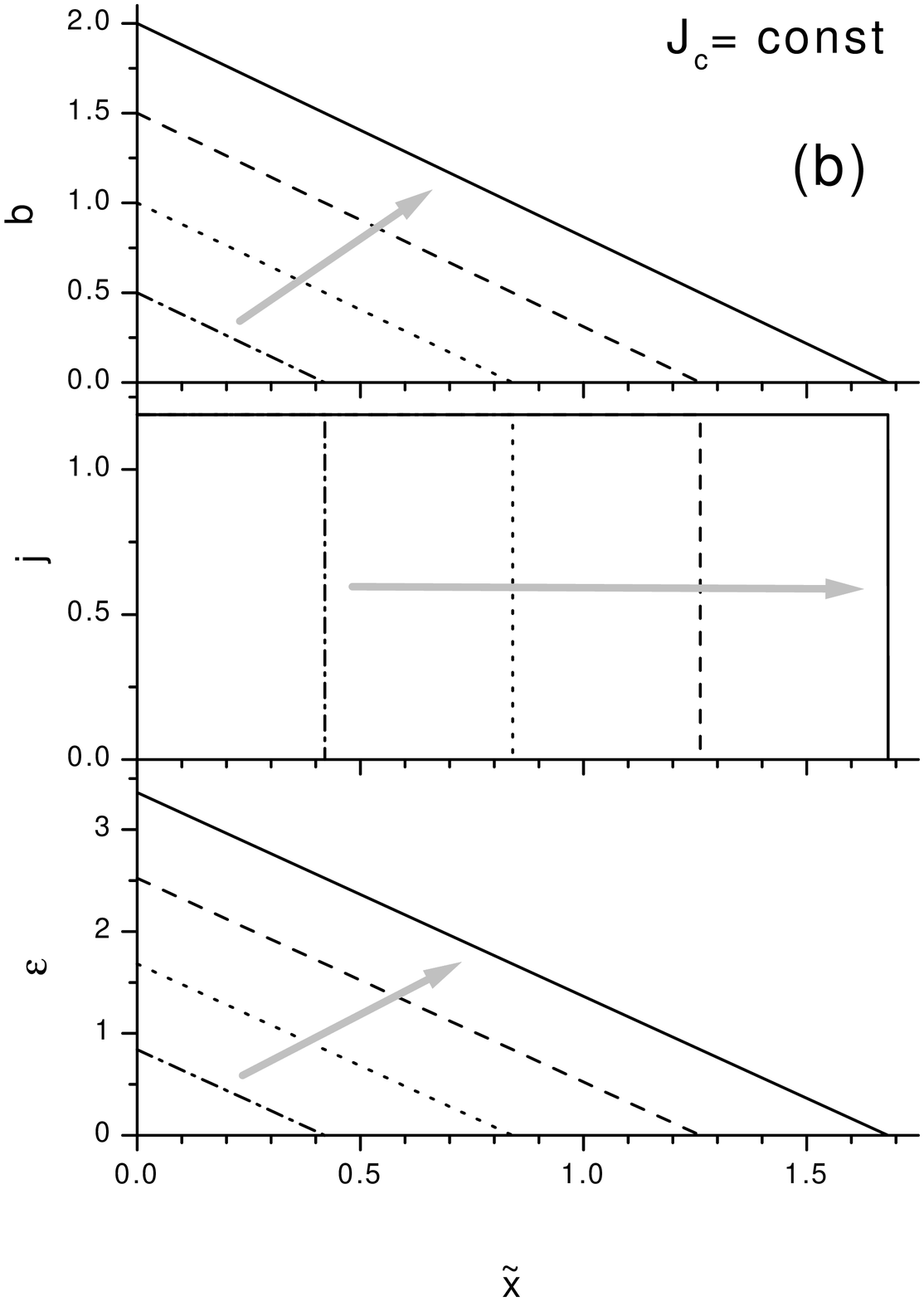,width=7.5cm}}
\centerline{\psfig{figure=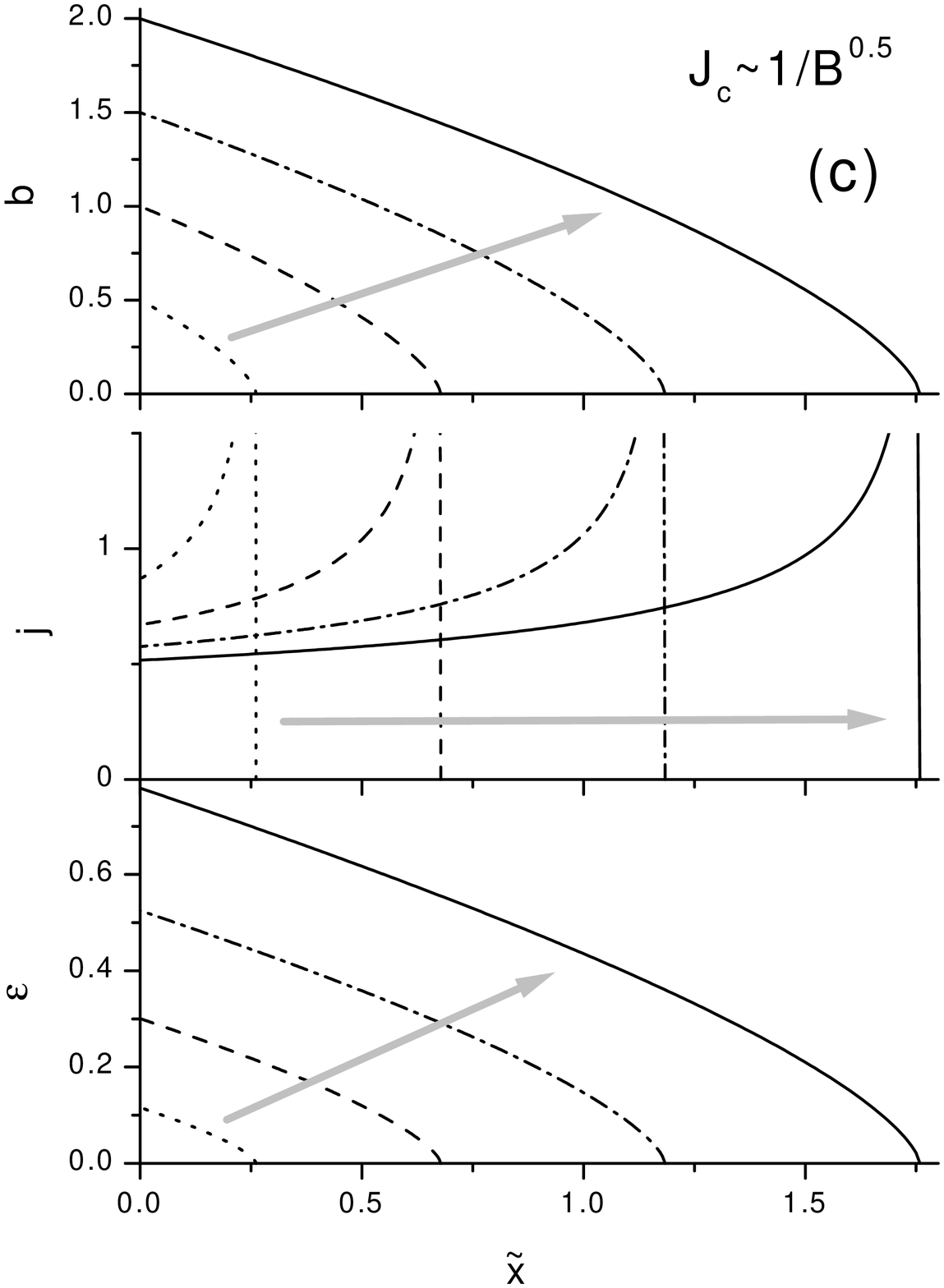,width=7.5cm}\psfig{figure=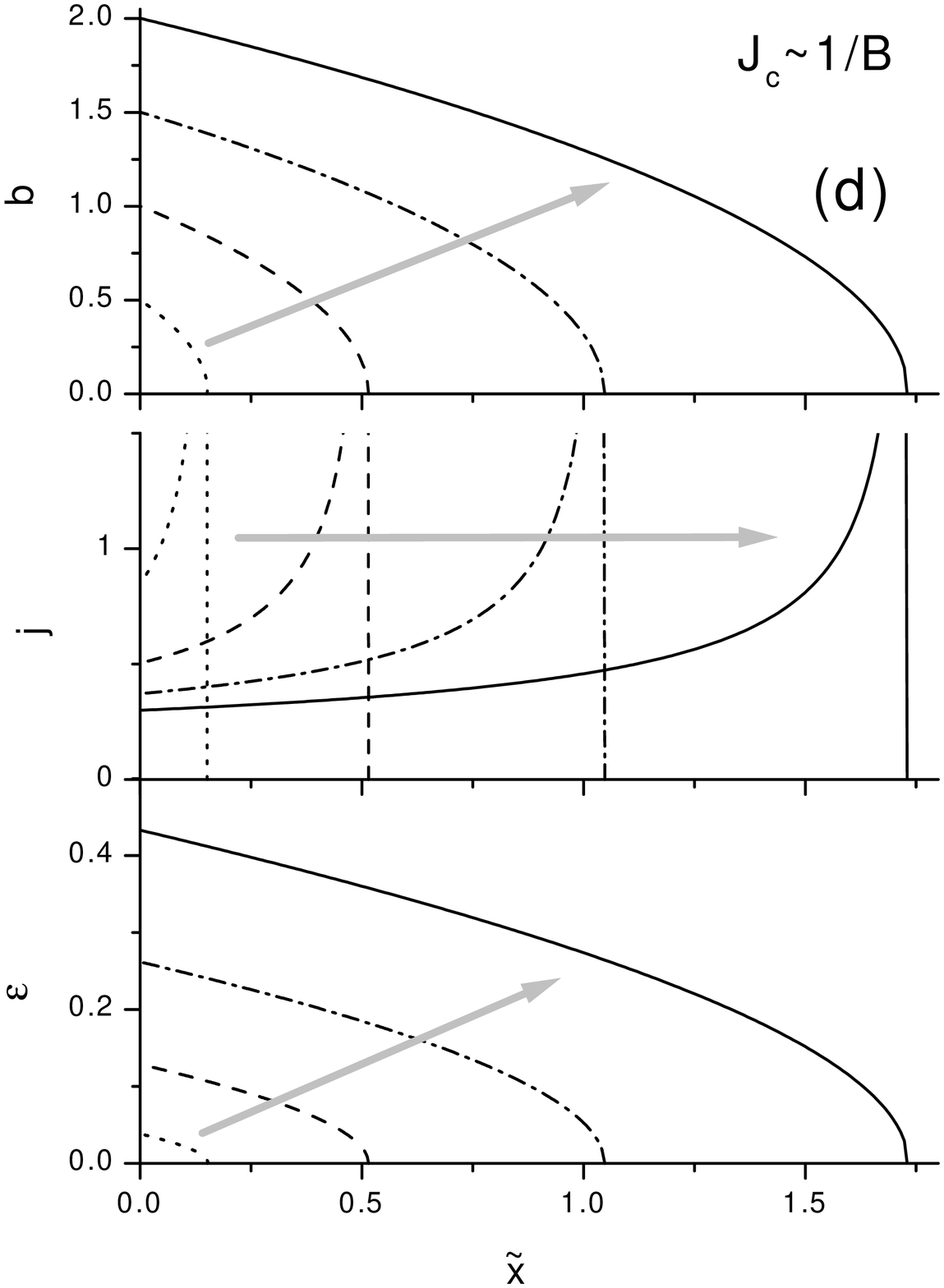,width=7.5cm}}
\caption{Distributions of the normalized flux density, current density
and electric field at equidistant times, $\tt=$0.5, 1, 1.5, and 2.
The applied field is {\em linearly increasing}, and $n=3$.
(a)-(d) differ in $J_c(B)$ dependence: 
(a)~$B$-independent $E(j)$, \eq{EJ};
(b)~constant $J_c$ ($J_c=2\dBadt / \mu_0 v_0$), a case 
solved exactly in Sec.~\ref{exactlin};
(c)~and (d)~$J_c$ decreases with increase of $|B|$.
The arrows indicate the time direction.
\label{f_a1}}
\end{figure}

The evolution of 
the flux density, current density, and electric field distributions
found numerically is presented in several graphs.
Figure~\ref{f_a1} shows results for the applied field 
linearly increasing in time, while \f{f_a0} as well as 
\f{f_exact2} -- for an instant turn-on of a constant field.  
{}From these graphs the following conclusions can be drawn.

(i) The flux density profile is found convex for  
$\gamma \ge 0$, i. e., for 
constant $J_c$ or $J_c$ monotonously decreasing with $B$.
It also means that $dJ/dx>0$, i.~e. the current density 
increases as $x$ approaches the flux front.
For $B$-independent $E(J)$ the picture is the opposite;
$|J|$ is maximum at the edge, and the $B(x)$ profile
is concave, see \f{f_exact2} and \f{f_a1}(a).
This behavior is expected for a field-cooled superconductor
when a small additional field is applied.
When comparing these results to experimental data one should keep in mind that
the shape of the observed $B(x)$ profile can easily be affected by 
demagnetization effects. 
However, the sign of $dJ/dx$ for the current density profile 
seems to be robust, which is confirmed by the flux creep simulations 
for a thin film geometry.\cite{schuster,brandt-sus97,mocur,creepsim}   
Schematic $J(x)$ profiles for the most important cases are shown 
in Tab.~1.

(ii) Even when $B_a$ increases linearly with time, the flux 
penetration can proceed with acceleration or deceleration  
depending on the $J_c(B)$ dependence.
This is illustrated in \f{f_a1} where the panels (a)-(d) are ordered 
according to the $J_c(B)$ behavior, and 
the time intervals between the curves are equal. Only in (b)
is the flux front moving with a constant velocity. 
In (a) the speed is slowing down, while in (c)~and (d) it is increasing. 
In the latter case, ramping of the applied field leads to an 
effective reduction of $J_c$, and hence to larger electric fields and 
faster penetration. (a), (b), and (c,d) correspond to $\beta<1$, $\beta=1$  
and $\beta>1$, respectively, where the exponent $\beta$ given by \eq{beta} controls the growth of the flux penetrated region, \eq{ff}.

\widetext
\begin{table}
\vbox{
\caption{The shape of the current density profiles and the 
parameter $\xi_0$ characterizing the flux front dynamics, \eq{ff},  
in different regimes. The parameter $a$ describes the approximate solution 
\eq{fapp} for the scaling function $f$.}

\begin{tabular}{c|c|c|c|c}
$n$  &  $B$-independent $E(j)$ & $j_c=$ const &$j_c(B)\propto 1/B$ &
The Kim model\\ 
   & \hspace{1cm}~($\gamma=-1/n$)\hspace{1cm}~ &
   \hspace{1cm}~($\gamma=0$) \hspace{1cm}  
   &  \hspace{1cm}~($\gamma=1$)  \hspace{1cm}~&
   \hspace{1cm}~$J_c=J_{c0} (1+B/kB_0)^{-1}$  \hspace{1cm}~\\ 
\hline
&\multicolumn{4}{c|}{} \\
&\multicolumn{4}{c|}{Constant field, $B_a = B_0 \Theta(t)$ \ \ ($\alpha=0$)} \\
&\multicolumn{4}{c|}{~} \\
\hline
 & \psfig{figure=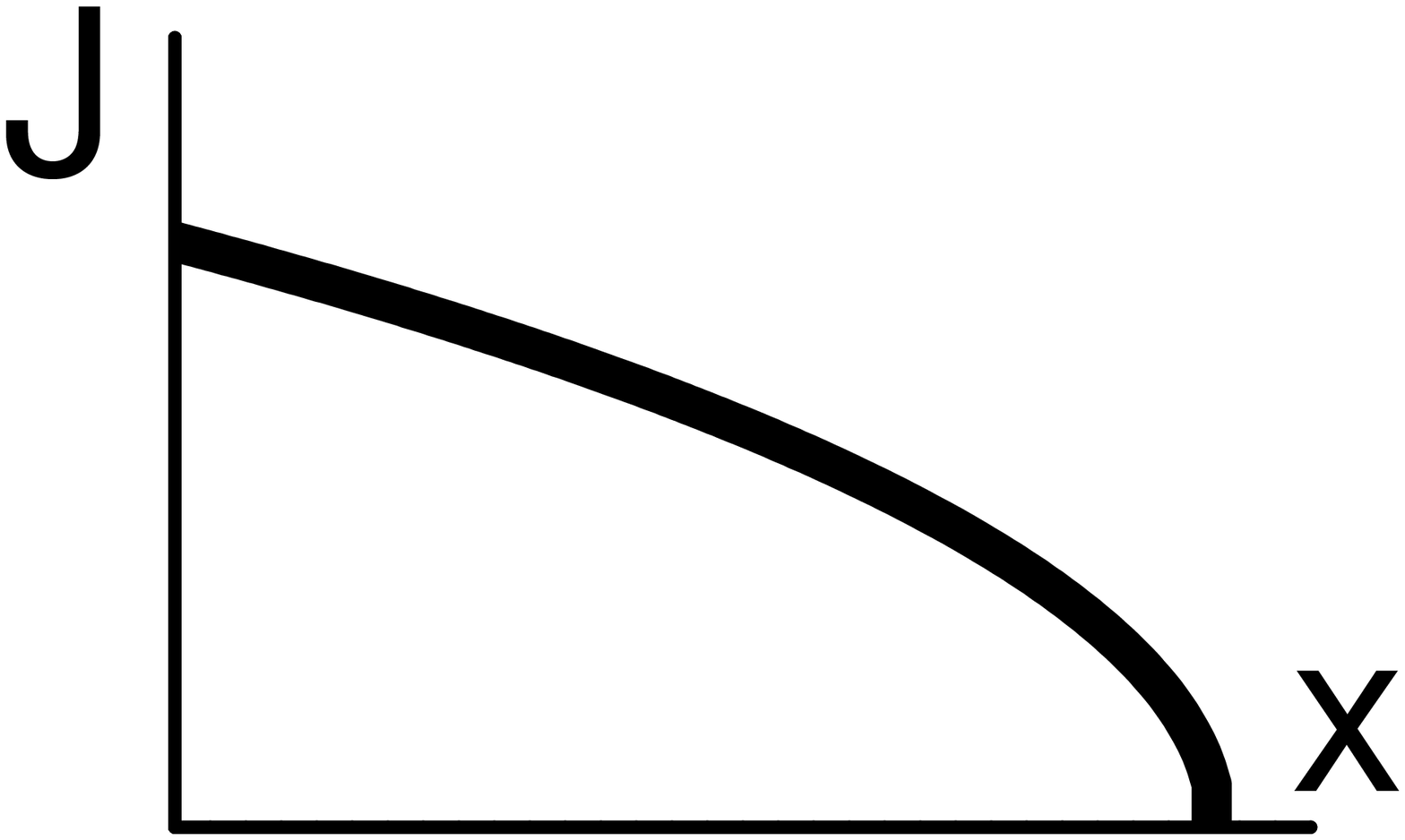,width=2cm}&  \psfig{figure=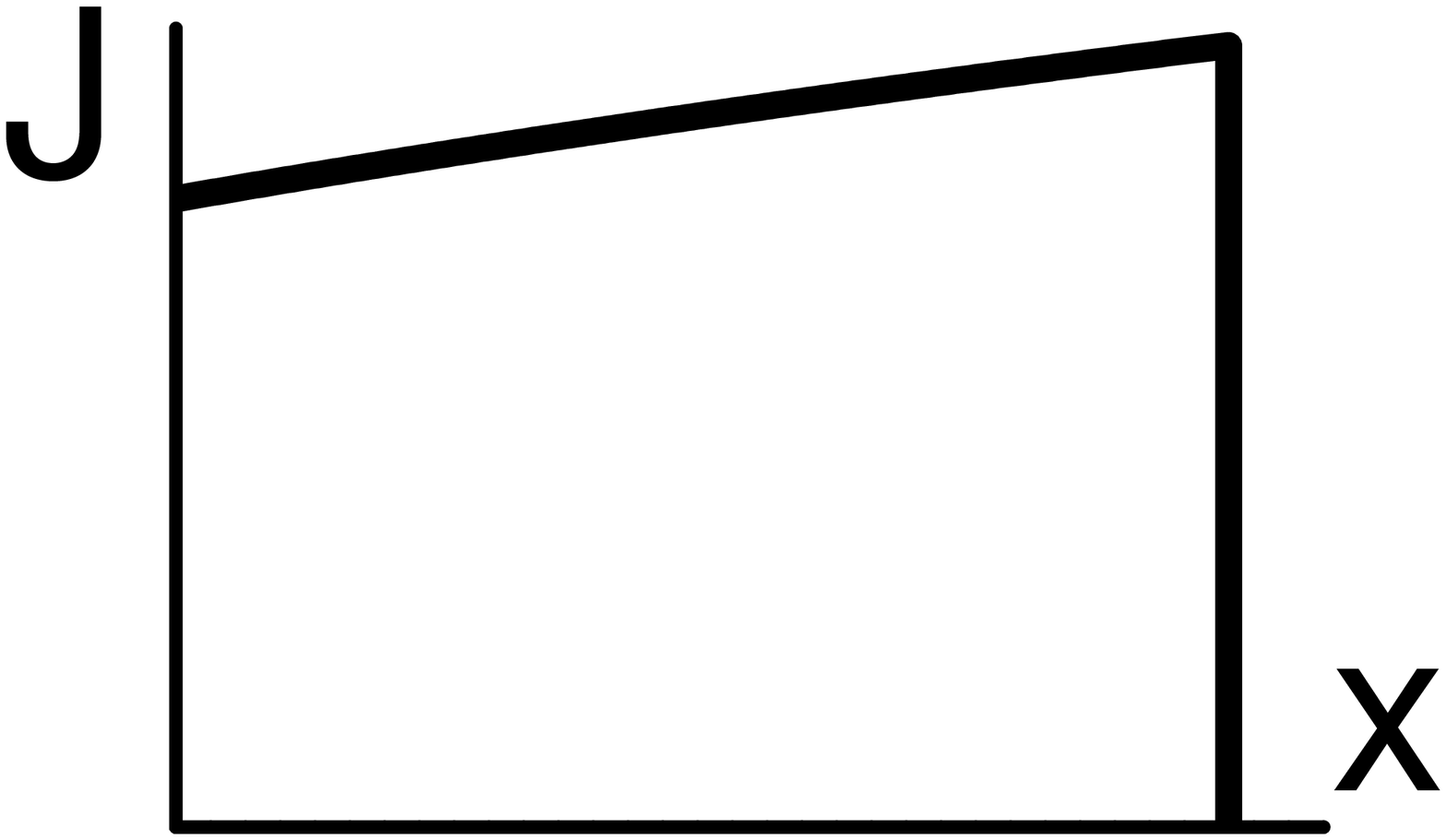,width=2cm} 
 & \psfig{figure=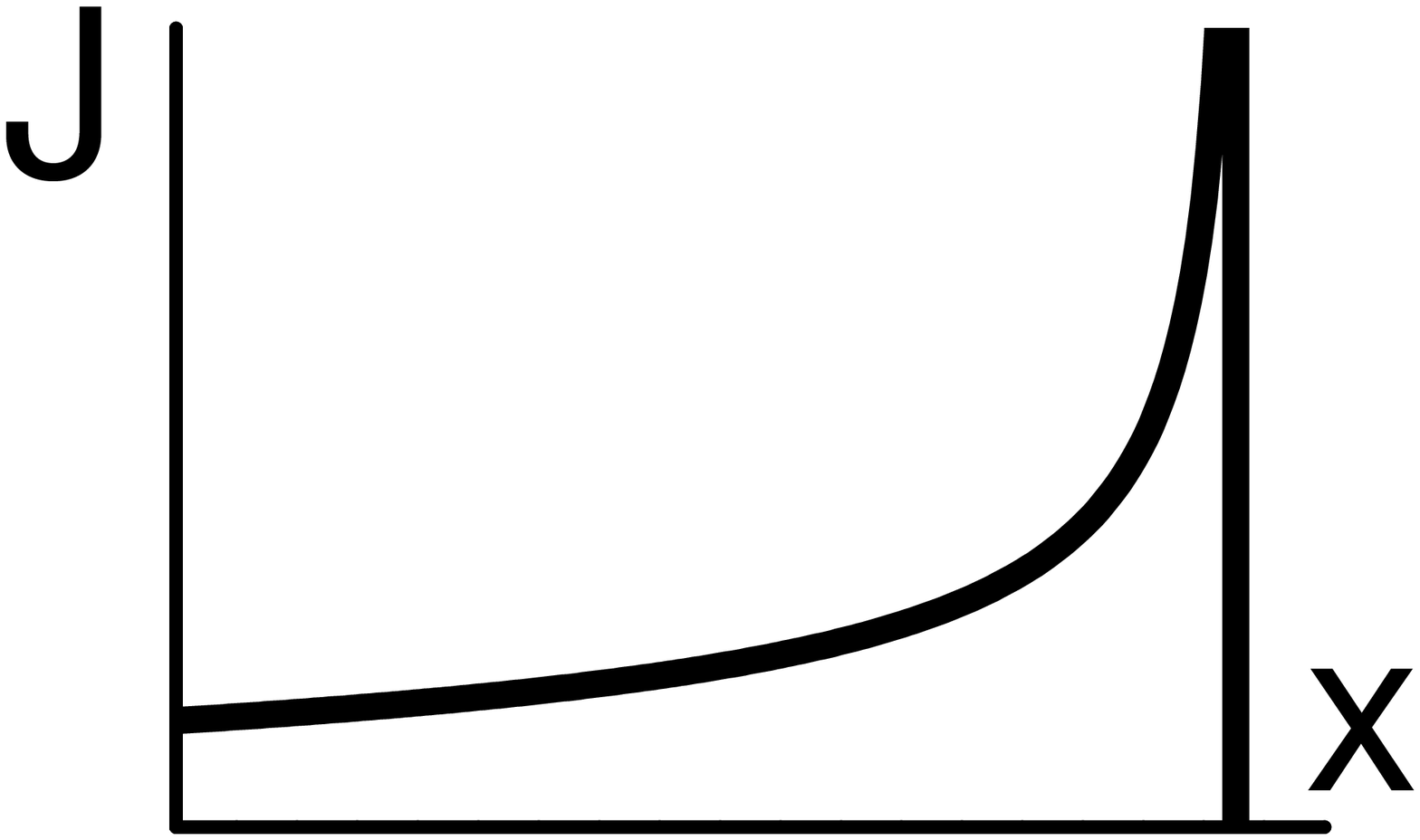,width=2cm}& \psfig{figure=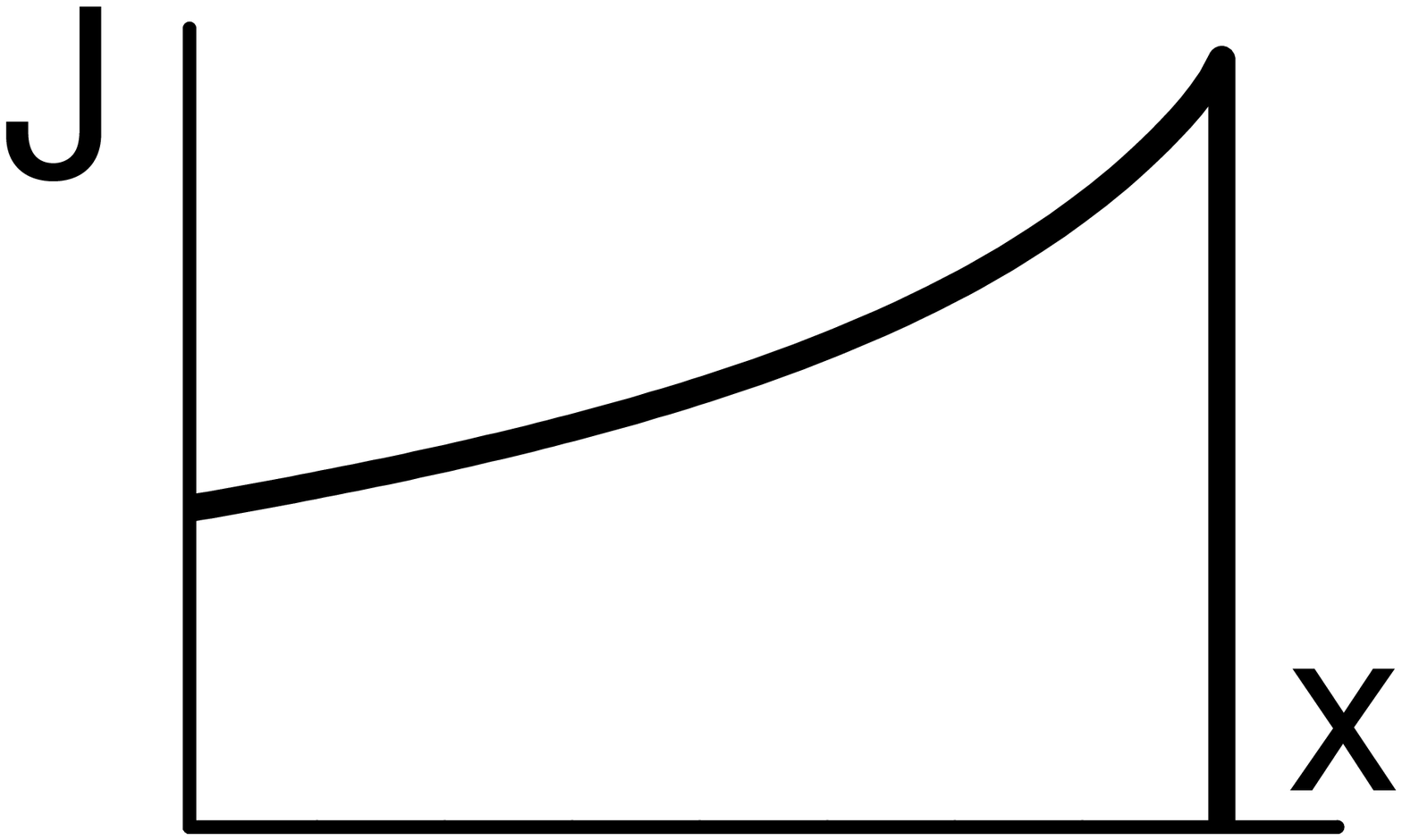,width=2cm}\\
 & see \f{f_exact2} & see \f{f_a0}(a) & see \f{f_a0}(b) & see \f{f_a0}(c,d)\\
\hline
 3 &  & $\xi_0=1.521$, $a=0.105$ & $\xi_0=0.881$, $a=0.032$ &  \\
\cline{1-1}
\cline{3-4}
7 & exact solution,& $\xi_0=1.346$, $a=0.045$ & $\xi_0=0.724, a=0.014$ &
$\xi_0=2.044$, $a=0.42$ for $k$=1, $n$=3\\ 
\cline{1-1}
\cline{3-4}
15 & \eq{bF}& $\xi_0=1.212$, $a=0.021$ & $\xi_0=0.628, a=0.007$ 
& $\xi_0=1.702$, $a=0.25$ for $k$=3, $n$=3  \\
\cline{1-1}
\cline{3-4}
33 & & $\xi_0=1.119$, $a=0.010$ & $\xi_0=0.568, a=0.004$ &\\
\hline
\hline
&\multicolumn{4}{c|}{~} \\
&\multicolumn{4}{c|}{Linear increase, $B_a = \dBadt \ t$ \ \ ($\alpha=1$)} \\
&\multicolumn{4}{c|}{~} \\
\hline
~ & \psfig{figure=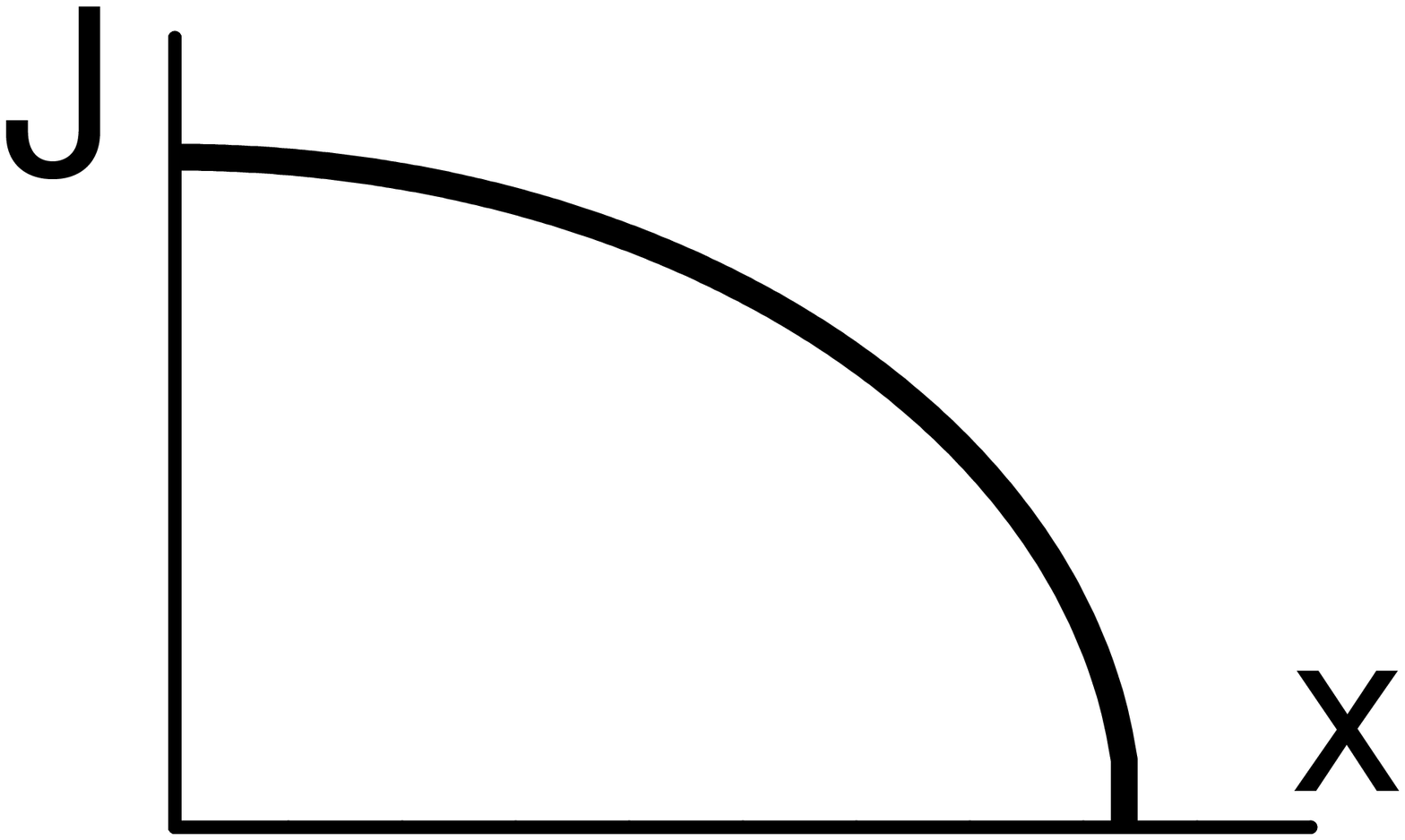,width=2cm}&  \psfig{figure=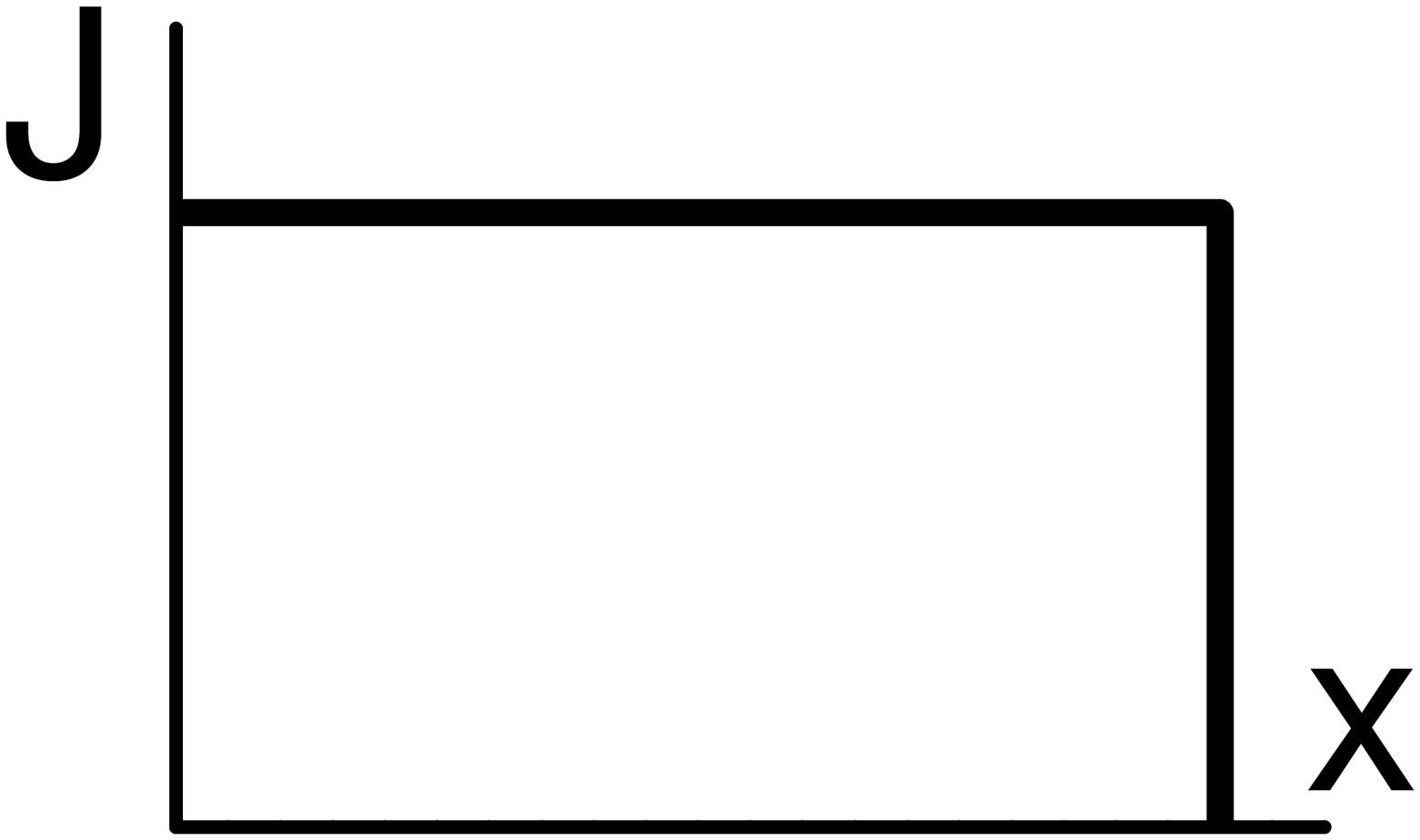,width=2cm}&
\psfig{figure=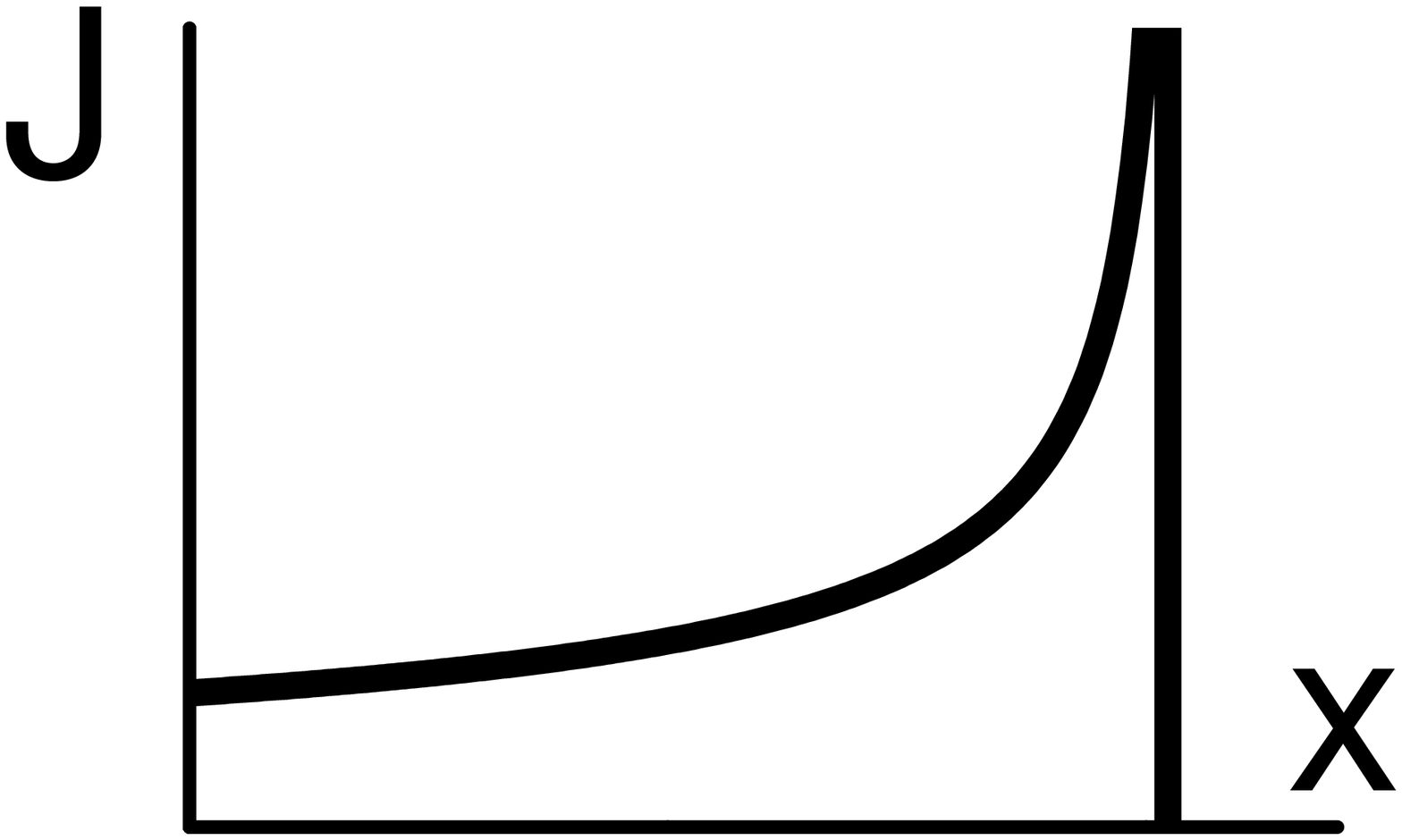,width=2cm} & NO\\
 & see \f{f_a1}(a) & see \f{f_a1}(b) & see \f{f_a0}(d)&SCALING\\
\cline{1-4}
 3  & $\xi_0=1.468$, $a=0.016$& & $\xi_0=0.514$, $a=-0.004$&\\
\cline{1-2}
\cline{4-4}
7  & $\xi_0=1.164$, $a=0$ & exact solution, & $\xi_0=0.503, a=-0.002$&\\
\cline{1-2}
\cline{4-4}
~~15~~ & $\xi_0=1.071$, $a=0$ & Eqs.~(\ref{t-x}-\ref{t-xx0}) & $\xi_0=0.501,
a=0.000$& \\
\cline{1-2}
\cline{4-4}
~~33~~ & $\xi_0=1.031$, $a=0$ & & $\xi_0=0.499, a=0.001$& \\
\end{tabular}
}
\end{table}
\narrowtext
  
\widetext

\begin{figure}
\centerline{\psfig{figure=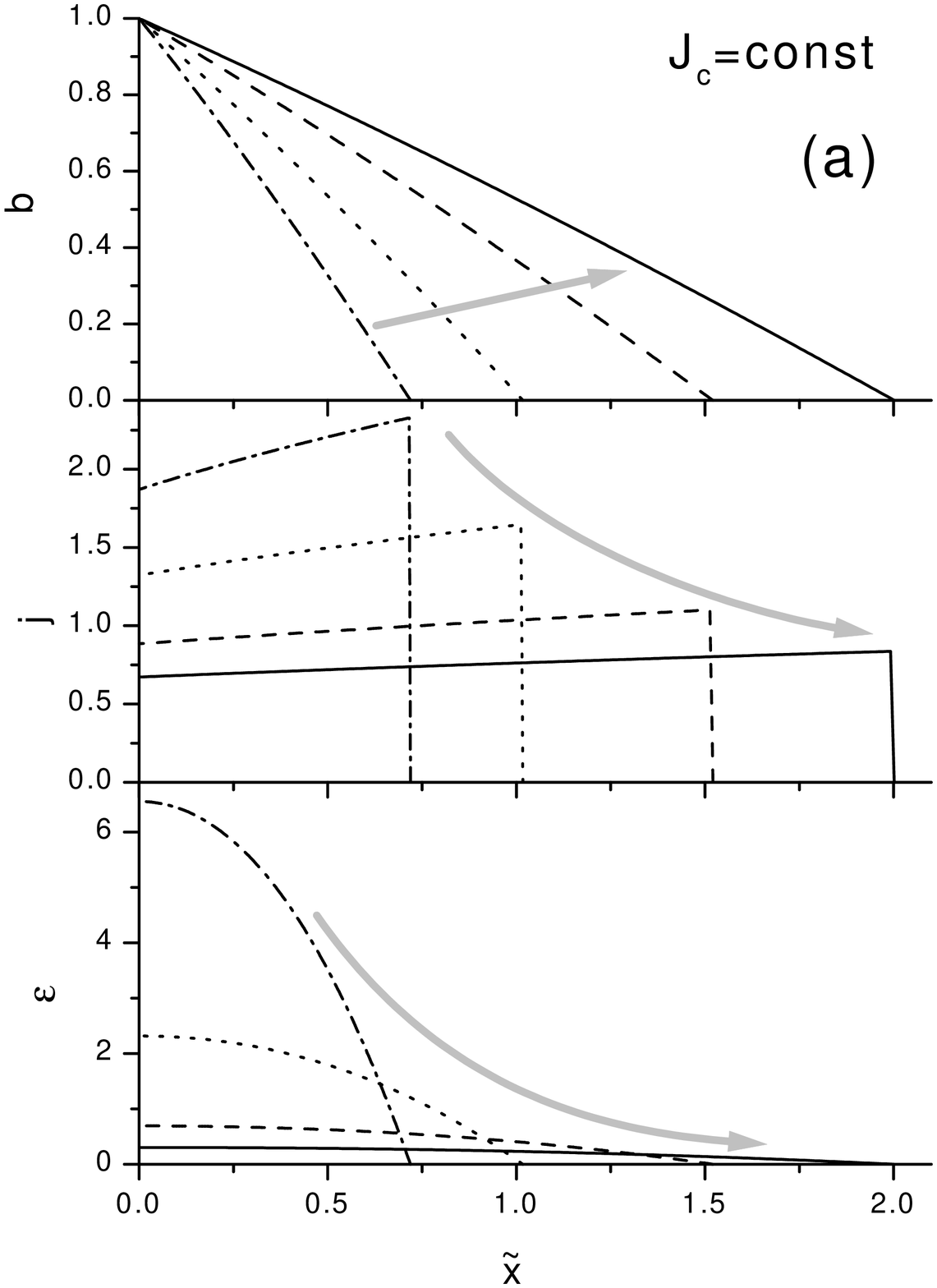,width=7.5cm}\psfig{figure=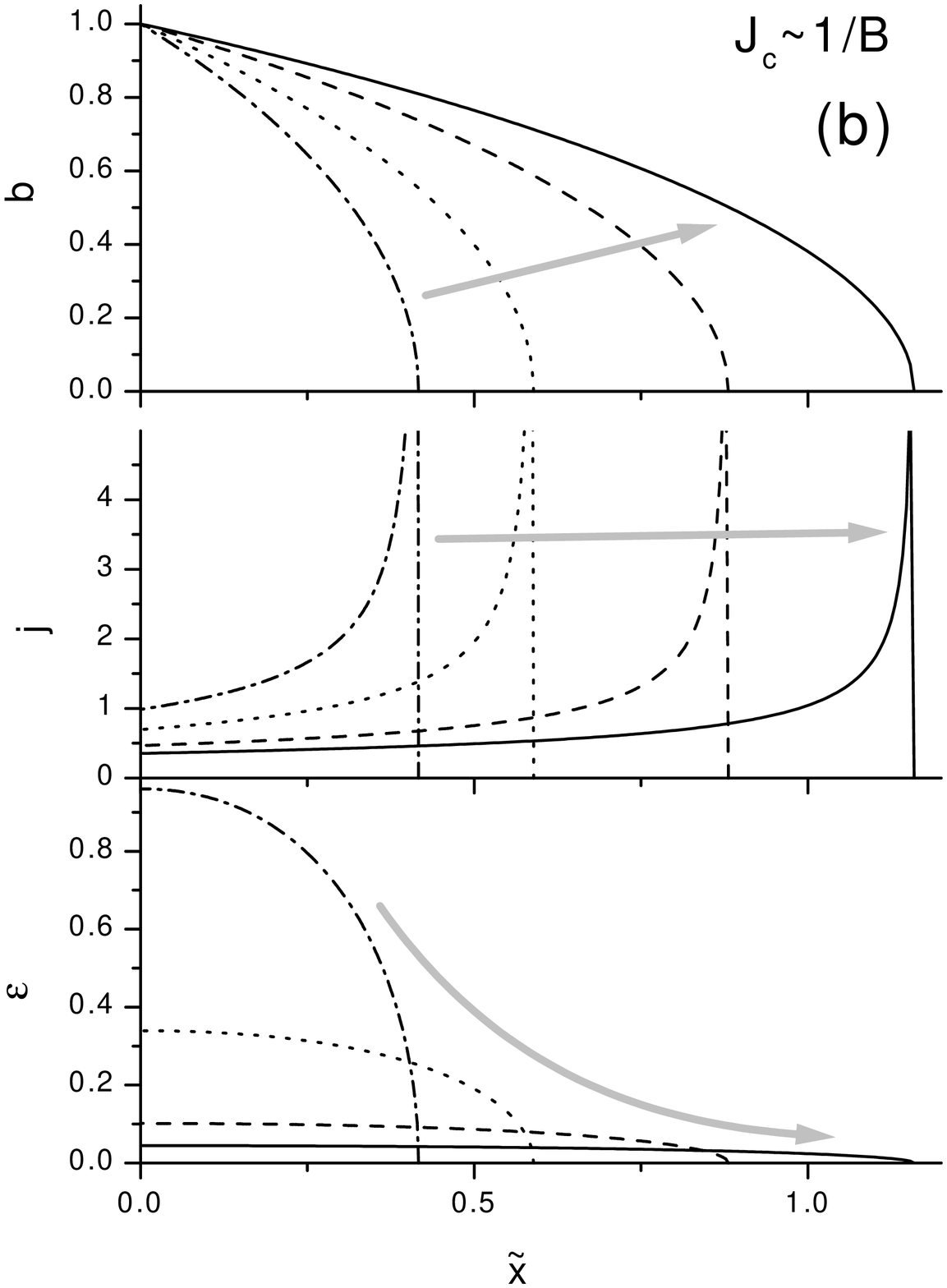,width=7.5cm}}
\centerline{\psfig{figure=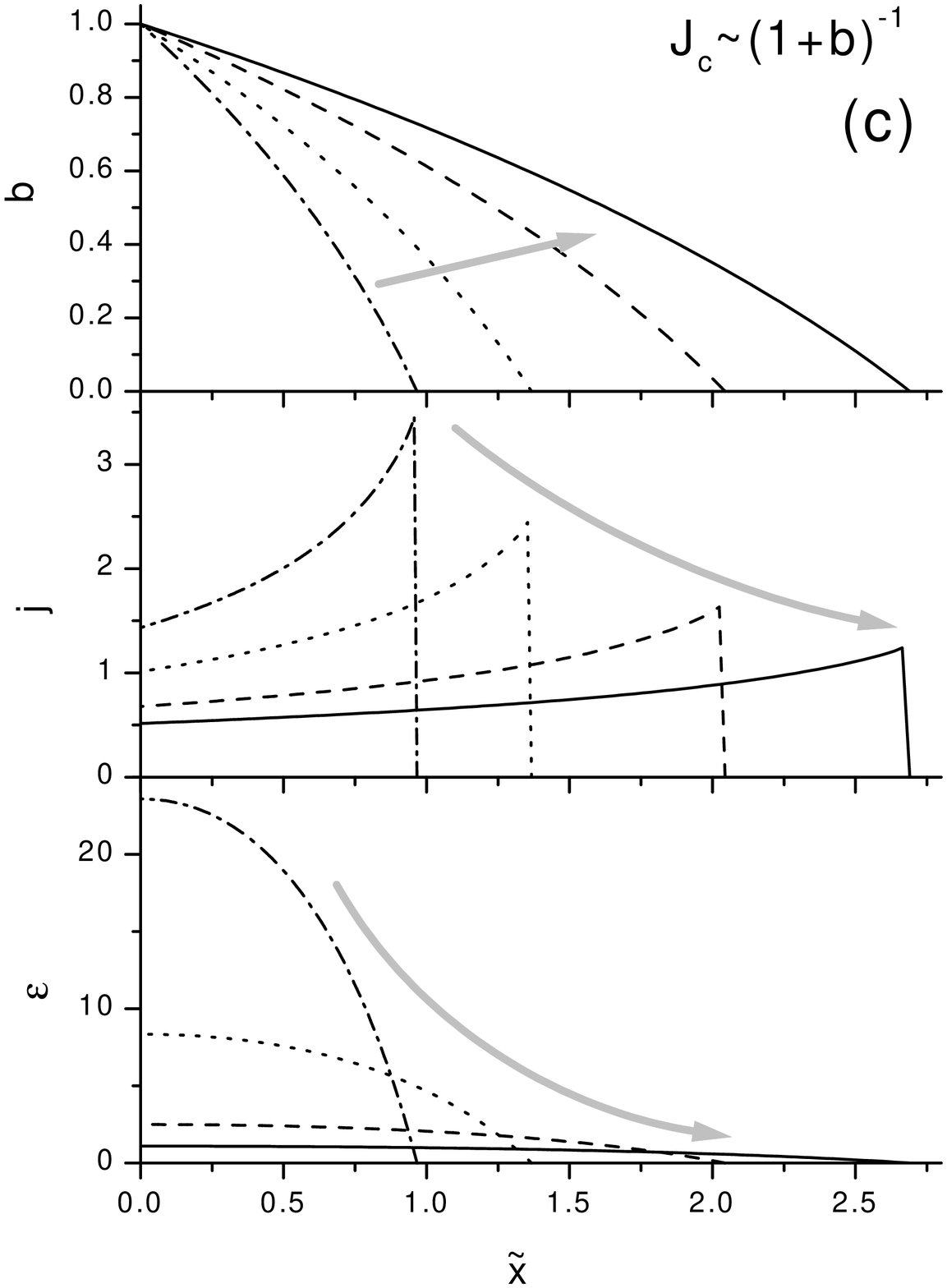,width=7.5cm}\psfig{figure=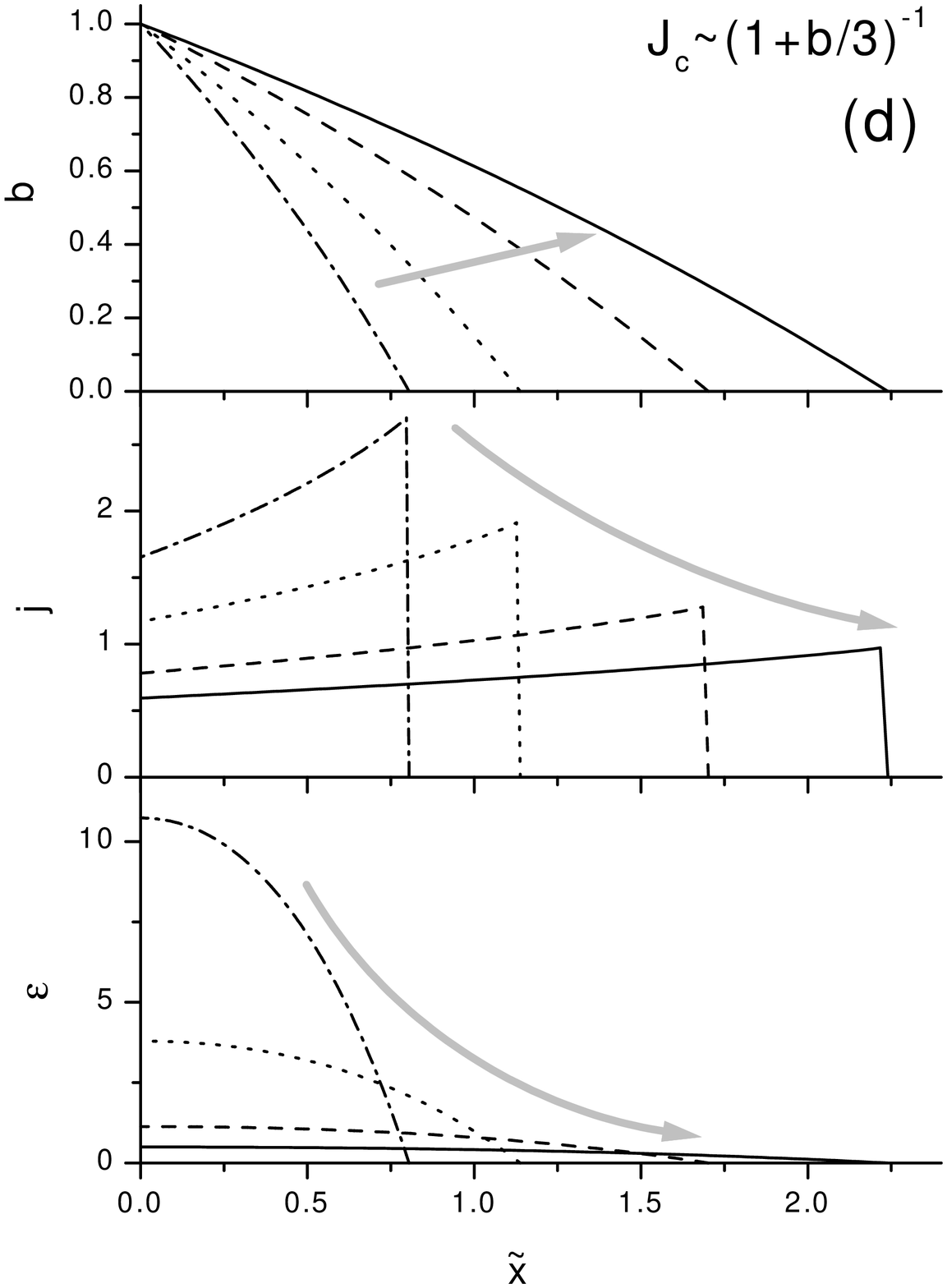,width=7.5cm}}
\caption{Distributions of the normalized flux density, current density
and electric field at different times, $\tt=$0.05, 0.2, 1, and 3.
The applied field is turned on at $\tt=0$ and kept {\em constant}, and $n=3$.
(a)-(d) differ in $J_c(B)$ dependence:
(a)~constant $J_c$; (b)~$J_c=J_{c0}/b$; (c)~and (d) the Kim model,
$J_c=J_{c0}/(1+b)$ and $J_c=J_{c0}/(1+b/3)$, respectively.
The arrows indicate the time direction. 
\label{f_a0}}
\end{figure}
\narrowtext

(iii) The electric field has always a maximum at the edge
and decreases monotonously with $x$, which follows from 
$\d E/ \d x = - \dot{B}\le 0$. Physically, this occurs simply 
because all the vortices enter the slab through the edge region where 
the flux motion is thus most intense.   
The second space derivative of $E$ is proportional to 
$\d J/\d t$, and one can find different curvatures of the $E$-field profile
for different cases. 

\section{Magnetization\label{s_mag}}

Since the nonlinear flux diffusion has a sharp front located a
finite distance from the surface, the solution applies also
to the case of a slab of {\em finite} width $2w$ provided $w$ is larger
than the penetration depth, $x_0$. The flux then penetrates  the slab from
both sides with non-overlapping profiles.

The time-dependent magnetization of
the slab is given by
\be
- \mu_0 M= B_a - \frac{1}{w} \ \int_0^w B(x)dx \, , 
\ee
where $B$ is non-zero only in the region, $0<x<x_0(t)$.
Substituting here the scaling law, \eq{sc}, one obtains
\beq
- \mu_0 M(t) = B_a(t) \, \left[ 1 -
     A\ \frac{x_0(t)}{w} \right] \, ,
\quad 0 \leq x_0 \leq w \, ,
\eeq 
where $A$ is a number equal to the average value of the 
scaling function in the penetrated region:
\be 
  A = \frac{1}{\xi_0}\ \int_0^{\xi_0} f(\xi)\ d\xi \, .
\ee 
For the exact solution considered
in Sec.~\ref{exactcon} one finds $A=(n-1)/[2nF(1)]<1/2$.
For the exact case in Sec.~\ref{exactlin} where the 
$B$-profile is linear, one has $A=1/2$. 
For flux density profiles with a convex shape, which are found when
$J_c$ decreases with $B$, one has $A>1/2$.

Expressing the time-dependence explicitly, the magnetization can
be written as 
\beq 
- \mu_0 M(\tilde{t}) = B_a(\tt) \, \left[ 1 -
      A\
(  \tilde{t}/\tilde{t}^*)^\beta \right]  \, , \quad 0 \leq \tilde{t} \leq
\tilde{t}^* \, ,
\label{Mt}
\eeq 
where $\tilde{t}^*$ is the time it takes for the flux to
completely penetrate  the slab, i.~e. $\tilde{t}^* =
(\tilde{w}/\xi_0)^\beta$.
For the fully-penetrated state, $\tt>\tt^*$, the time-dependence of $M$
is essentially different, and one expects a kink in the magnetization
relaxation rate, as first predicted in Ref.~\onlinecite{vinokur} 

\section{Transport current and non-stationary $V(I)$\label{s_v}}

Most results found in the paper are also relevant to 
a superconductor carrying a transport current.
The flux penetration into a 
slab biased with transport current is governed by the same equations
as those for a slab placed in an applied magnetic field. 
The difference is that for the transport current case
the current flows in the same direction
on both edges of the slab.
However, this is not important as long as the two penetrating 
flux fronts do not meet in the slab center.
The boundary condition $B(x=0)=B_a$ for the applied field case
should be replaced now by $B(x=0)=\mu_0 \IT/2$, where $\IT$ is the transport
current
per unit height of the slab.

It is interesting to analyze the voltage $V$ measured on the
superconductor carrying a transport current $\IT$.
It is obvious that $V$ is not a unique function of $\IT$
but also strongly depends on how $\IT$ changes with time.
In a conventional experimental setup the voltage contacts are
attached to the superconductor surface, thus
the measured voltage is determined by the electric field
in the surface layer only, $V(t)=E(x=0,t) L$, where $L$ is the distance
along $y$ between the contacts. 

The electric field at the edge has a power-dependence on time given by
\eq{jesc}, since $f_\epsilon(0)$ is just a number.
Thus, from \eq{beta} one obtains the following power-law 
voltage-current relation,
\be
  V \propto \IT^p, \quad p=1+\frac{n}{n+1} \l( 1-\frac1\alpha+\gamma\r).
\label{VIpower}  
\ee
Note that the exponent $p$ of the integral $V(I)$ can be much different from the
exponent $n$ characterizing the local $E(j)$. Actually, $p$ is only weakly
sensitive to $n$, especially for large $n$. Moreover, 
{\em increase} in $n$ can sometimes lead to a {\em decrease} in $p$.

For large transport currents when the flux penetrates the whole sample
the present analysis is not valid. Then   
the current density will be distributed over the
slab more or less uniformly, and one can expect that
the integral voltage-current curve will reflect the local one, 
i.~e., $V \propto J^{n+1}$. 
The crossover between low-current and high-current parts in $V(I)$
has been reproduced by numerical simulations in Ref.~\onlinecite{zhang}.
This crossover should be accessible experimentally since 
the exponent $p$ for an incomplete flux penetration  
is of the order of unity, 
while the exponent $n$ for the full penetration is
temperature-dependent and can be very large.

The exponent $\alpha$ in \eq{VIpower} defines the time dependence of the transport
current, $\IT(t) \propto t^\alpha$. 
For larger $\alpha$
$V(\IT)$ becomes steeper until the exponent $p$ saturates. 
For $\alpha=0$, i.~e. when the current is turned on and kept constant, the
voltage decays with time.  
This voltage relaxation has been earlier observed
experimentally\cite{zhang99,ma,zeng} 
and reproduced by numerical simulations.\cite{zhang99,zeng}   
It is accompanied by relaxation of current
distribution, as illustrated by \f{f_a0}. 
According to \eq{jesc}, the voltage decay is described by 
$V \propto t^{-n/(n+1)}$, in particular 
$V \propto 1/t$ at low temperatures when $\nn$ is large.

Remarkably, for small enough $\alpha$
the voltage will {\em decrease} with {\em increasing} current. 
For example, for $\alpha=1/3$ and $\gamma=0$, one obtains
\be
V \propto \IT^{-(n-1)/(n+1)},
\label{crazyVI}
\ee
in particular, $V \propto 1/\IT$ at low temperatures.
Physically, a decrease of voltage is related to the same relaxation process
which takes place for any external conditions.
Increase of $\IT$ with time supplies more current to the superconductor
and tends to increase the voltage. If $\IT$ increases slow enough,
then the first process is dominant, and the measured $V(\IT)$ curve
should give voltage decreasing as the current increases. 
The evolution of flux, current and electric field distributions for this
interesting case is shown in \f{f_crazyVI}. 
\begin{figure} 
\vbox{
\centerline{\psfig{figure=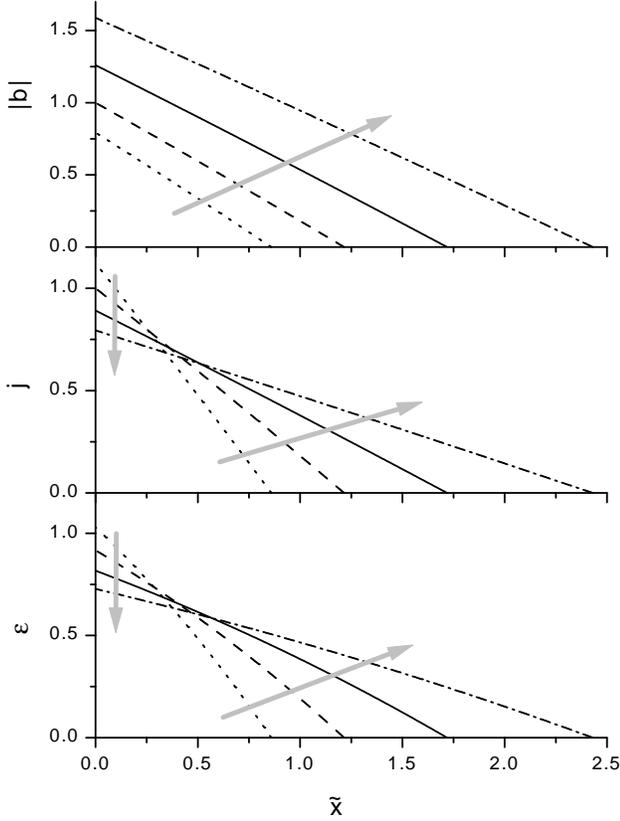,width=8.5cm}}
\caption{Distributions of the normalized flux density, current density
and electric field at different times, $\tt=$0.5, 1, 2, and 4.
The transport current increases with time as $\IT\propto t^{1/3}$, while 
$J_c=const$ and $n=3$. The voltage, which is proportional to $\epsilon(x=0)$,
decreases with time, as shown by the arrows indicating 
the time direction. 
\label{f_crazyVI}}}
\end{figure}

For the experimentally most relevant case of a linearly increasing
current, $\IT=\dIdt\, t$,  (i.e. $\alpha=1$) one obtains from \eqs{jesc}{norm}
\be
 V = \frac{\mu_0 v_0 L|f'(0)|^n }{2}\ 
 \l(\frac{\mu_0^\gamma \dIdt}{2^{1+\gamma} v_0 J_{c0}
 B_0^\gamma}\r)^{\frac{n}{n+1}} 
     \, \IT^{1+\frac{\gamma n}{n+1}} .
\label{vic}       
\ee
For a constant $J_c$, or $\gamma=0$, we come to
the exactly solvable case considered in Sec.~\ref{exactlin}, and 
the expression further simplifies to,  
\be
  V = \frac{\mu_0 v_0 L}{2} \l(\frac{\dIdt}{2 J_c v_0}\r)^{n/(n+1)}\,
  \IT \, .
\label{vexactlin}  
\ee
In this case the superconductor behaves like an {\em Ohmic} conductor.
Deviations from the Ohmic behavior can be caused by a  
$B$-dependence of $J_c$. In particular, 
if $J_c$ decreases with $B$, the exponent of the $V(\IT)$ curve 
becomes larger than unity, see \eq{vic}.
Meanwhile, at small $I$ the superconductor always behaves Ohmically
since the self-field is small and the $J_c(B)$
dependence can be ignored. These results are in agreement with
numerical simulations reported in Ref.~\onlinecite{zhang-ic}.\cite{2-1}   
 
We also note that when the transport current is ramped faster, the
voltage at a given current is larger, $V \propto \dIdt^{~n/(n+1)}$. 
This observation is in agreement with results of numerical simulations
and experiment on Bi-based tapes.\cite{zhang99}

\section{Approximate solution}

In a general case \eqs{f}{fkim} with boundary conditions (\ref{boundary})
cannot be solved analytically.
Near the flux front $f \to 0$, and  \eq{f} reduces to $\l({\rm const}+\beta\xi_0\r)^{1/n} = f^\gamma\, |f'| $.
Therefore, the behavior of the scaling function $f$ 
is determined only by the $J_c(B)$ law at small $B$, and
\be
  f(\xi) \propto \l( \xi_0 - \xi \r)^{1/(1+\gamma)}, \quad \xi
  \rightarrow \xi_0 \, .
\ee
This result holds true for any $J_c(B)$ which has asymptotic behavior           
$J_c \propto B^{-\gamma}$ at $B \to 0$, e.g., $\gamma=0$ for the Kim model.

Surprisingly, a very good approximate solution for $f$ in the {\em whole} region $0 \le
\xi \le \xi_0$ 
is given by the expression
\be
  f(\xi) = \l( 1 - \xi/\xi_0 \r)^{1/(1+\gamma)} (1+a \xi)
\label{fapp}
\ee
Values of $\xi_0$  and $a$ for several common dependences
$J_c(B)$, including the Kim model, and values of $n$ 
have been found numerically and listed in Tab.~1. 
Expressions for $\xi_0$  and $a$ in a general case can be found by 
substituting $f$ from \eq{fapp} into \eq{f} and analyzing expansion in powers of
$(\xi_0-\xi)$.
Then, one obtains:
\begin{eqnarray}
   &&\frac{1}{a} = \frac{2n(1+p)(p+\alpha n)}{p^2 (p-\alpha)}  -1,
   \quad p=1/(1+\gamma) \, , \\
   &&\xi_0^{n+1} = \frac{n+1}{1+\alpha n(1+\gamma)}\,
   \frac{(1+a)^{n(1+\gamma)}}{(1+\gamma)^n}\, .
\end{eqnarray}
Since the scaling function $f(\xi)$ usually has a very simple shape,
we find that the approximate expressions fit the exact
solutions with a good accuracy, e. g., the deviation is less than 1\%
for all cases shown in Figs.~\ref{f_a1} and~\ref{f_a0}.

\section{Conclusions}

The propagation of magnetic flux into a slab superconductor
has been considered using the flux creep approach with a 
logarithmic current dependence of the activation energy. 
The dynamic behavior was found to possess
scaling for a general $J_c(B)$ when a constant magnetic field is suddenly applied to the superconductor, and for a power-law $J_c(B)$ 
in case of an applied field ramped up with a general 
power-dependence on time. 
For two particular cases of the creep problem an exact analytical solution
 could be found. 

The main results obtained in this work are: 
\begin{enumerate}
\item
The flux density profiles at different times follow the scaling
law, $B(x,t) = B_a(t) f(xt^{-\beta})$.
Similar scaling applies to the current density and electric field profiles.

\item
 The flux density profile is convex for penetration into 
zero-field-cooled slab, and concave for a slab cooled in a large field.

\item At constant $J_c$ and linearly increasing $B_a$ the $B(x)$
and $J(x)$ profiles
at any time coincide with the Bean-model profiles.

\item
The flux front position is a power function of time given by \eqs{ff}{beta}.
The front moves through the slab with an increasing or 
decreasing velocity depending on the material's $J_c(B)$. 

\item The explicit time dependence of the magnetization is found.

\item
For a partially penetrated slab carrying a transport current $I$, 
the voltage $V$ is a power of $I$, with an exponent different from that of
the local $E(J)$ relation. A pronounced crossover in the $V(I)$ curve
at the point of full penetration is predicted.

\item
 For a small transport current increasing linearly with time, the Ohmic behavior
 $V \propto I$, is found.

\item For a stationary transport current the voltage decays as $V \sim 1/t$.
 
\item An increase of transport current can be accompanied by 
a decrease of voltage, in particular, $V \sim 1/I$ when $I\propto t^{1/3}$. 
\end{enumerate}

All the conclusions can be tested experimentally:
(1)-(4) by spatially-resolved techniques, and 
(5)-(9) by integral measurements.
Our results presented by \eqs{Mt}{VIpower}  
allow a new method to infer the material 
properties such as local $E(J)$ or $J_c(B)$ characteristics
from integral measurements of magnetization and voltage.

\acknowledgements
The financial support  by  the Research Council of Norway
is gratefully acknowledged.

\widetext

\begin{references}
\bibitem{2param} Only a combination of $B_0$ and $J_{c0}$ enters the equation.
Hence $B_0$ can be chosen arbitrarily, which will be done below.
\bibitem{specialcase} For the special case $\alpha = 1/(1+\gamma)$, one cannot use
\eq{tau},
however we again come to the same \eq{b} only with an additional factor
$K =  {\mu_0 J_{c0} v_0 \tau}/{B_0}$
at the right-hand side of the equation.
\bibitem{2-1} For the average electric field calculated in 
Ref.~\onlinecite{zhang-ic} the scaling law is $\int E(x) dx \propto I^{p+\beta}$.
For $\alpha=1$, and $\gamma=-1/n$,  this exponent equals to $ 1+(n-1)/(n+1) \sim 2$.
\end{references}
\end{document}